\documentclass[aps,prb,reprint,longbibliography]{revtex4-2}
\usepackage{graphicx}% Include figure files
\usepackage{dcolumn}% Align table columns on decimal point
\usepackage{bm}% bold math
\usepackage{hyperref}% add hypertext capabilities
\usepackage{amsmath}
\usepackage{amssymb}
\usepackage{braket}
\usepackage{mathtools}
\usepackage{array}
\usepackage{float}
\usepackage[T1]{fontenc}

\begin{document}
\title{Qubit encodings in the p-orbital-valley spectrum for enhanced coherence and tunable two-qubit interaction}
\author{John H.~Caporaletti}
\affiliation{Department of Physics, University of Maryland Baltimore County, Baltimore, MD 21250, USA}
\author{J.~P.~Kestner}
\affiliation{Department of Physics, University of Maryland Baltimore County, Baltimore, MD 21250, USA}

\begin{abstract}
We propose encoding a qubit in a two-level subspace spanned by the lowest $p$-orbital state in the excited valley of an anisotropic quantum dot and the excited $p$-orbital in the ground valley, which we dub the $pOv$ qubit. There is an avoided crossing between these states due to valley-orbit coupling (VOC) induced by alloy disorder, enabling complete single-qubit control using baseband electrical control of the dot anisotropy. We find that `sweet spots' exist at specific dot orientations where the instantaneous eigenstates are first-order insensitive to charge noise. Using a phenomenological two-level fluctuator (TLF) dipole noise model, we estimate an average dephasing time of $T_2^*\approx 10\,\mu\text{s}$ and a quality factor of $Q\sim 10^4$. Alternatively, encoding in the $p$-orbital states in the ground valley near the isotropic dot point, we show that one can induce a similar sweet spot via an out-of-plane magnetic field. Finally, we find that two-qubit gates for the $pOv$ qubit are mediated by the quadrupole-quadrupole Coulomb interaction and can be electrically tuned from zero to $\sim 1~\text{GHz}$ by adjusting the relative orientation of the anisotropic dots, providing a novel pathway towards scalable quantum computation.
\end{abstract}

\maketitle
\section{Introduction}
Electrons confined to semiconductor quantum dots are a leading platform for quantum computing, offering compatibility with industrial fabrication processes, long coherence times, and the potential for large-scale integration~\cite{burkard_SemiconductorSpinQubits_2023}. Encoding a qubit in the charge degree of freedom (DOF) offers fast baseband control and readout, but couples strongly to electric field fluctuations through its dipole moment, resulting in fast decoherence $T_2^*\lesssim 1~\text{ns}$~\cite{gorman_ChargeQubitOperationIsolated_2005,petersson_QuantumCoherenceOneElectron_2010}. As a result, semiconductor qubits are typically encoded in the spin DOF. However, spin qubits are typically controlled resonantly (the triple-dot exchange-only qubit being a notable exception \cite{divincenzo_UniversalQuantumComputation_2000}), requiring oscillating magnetic or electric fields, producing slow operations and the possibility of device heating and scalability concerns.

One strategy for improving charge qubit coherence without abandoning baseband electrical control is to engineer qubit states with vanishing dipole moments. The charge quadrupole qubit~\cite{friesen_DecoherencefreeSubspaceCharge_2017} couples to the electric field through a quadrupole moment, suppressing sensitivity to spatially uniform electric field fluctuations. This concept was experimentally demonstrated using a single electron in a triple quantum dot~\cite{koski_StrongPhotonCoupling_2020}, where strong coupling to the quadrupole moment was achieved through a superconducting resonator. However, the triple-dot quadrupole qubit introduces a low-lying leakage state, is sensitive to dipolar noise couplings past second order, and requires three dots per qubit, increasing the device footprint.

A recent quadrupole qubit proposal~\cite{caporaletti_ProposedFiveElectronCharge_2025}, dubbed the $pO$ qubit, instead encodes in the two-dimensional $p$-orbital subspace of the valence electron of a five-electron silicon quantum dot. The $pO$ qubit states have no dipole moment at any operating point so the leading-order coupling to charge noise is always through the quadrupole moment. This encoding has several attractive features: it uses a single quantum dot, is compatible with existing spin qubit devices, has no low-lying leakage states, features all-electrical control through modulation of the dot eccentricity and/or orientation, and supports fast two-qubit gates via the quadrupole-quadrupole Coulomb interaction. Rabi frequencies can easily be on the order of GHz, and the $pO$ qubit was estimated to yield an order-of-magnitude improvement in quality factor relative to state-of-the-art semiconductor spin qubits.

\begin{figure}[t]
    \centering
    \includegraphics[width=.65\columnwidth]{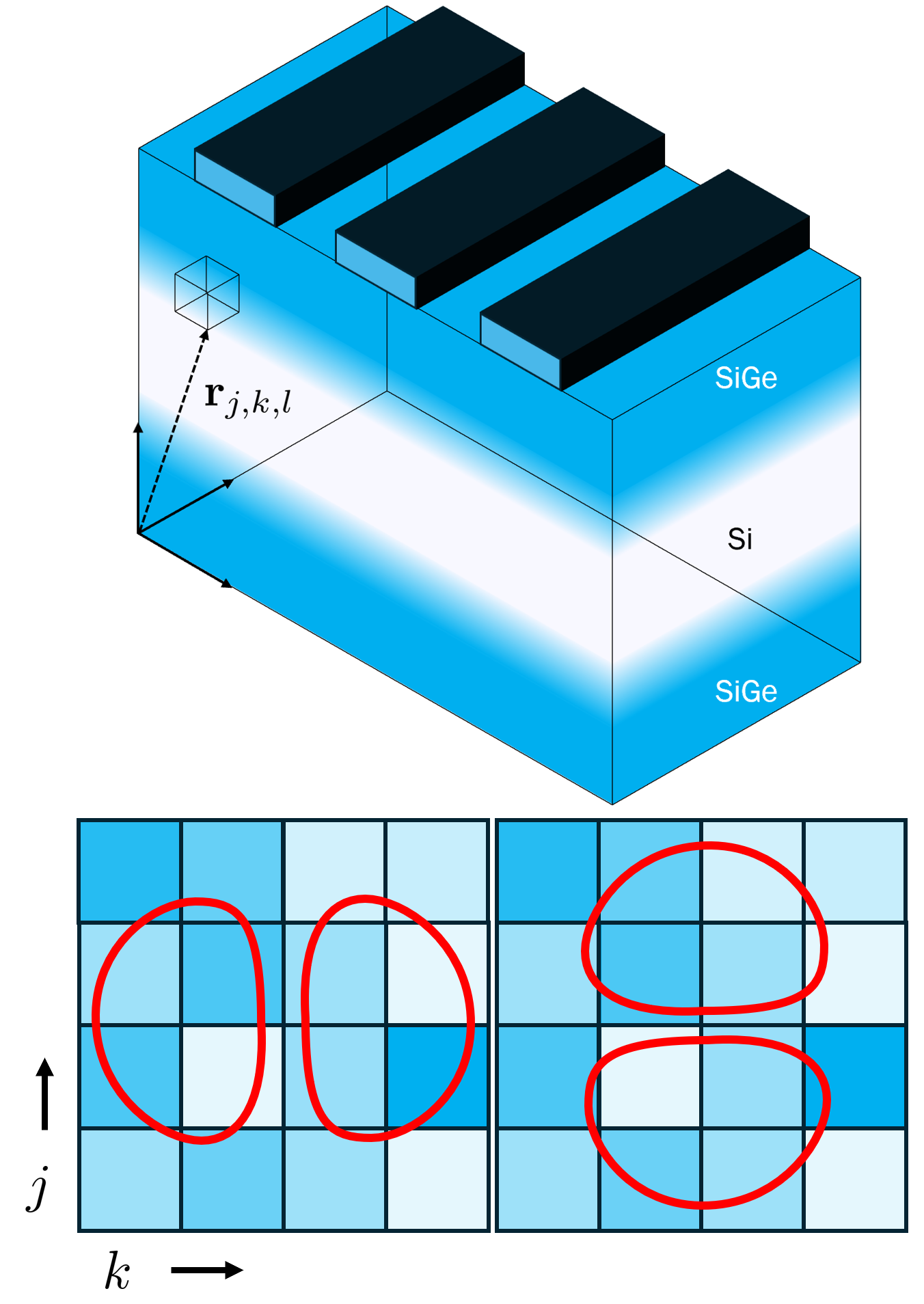}
    \caption{Top: cartoon depiction of SiGe/Si/SiGe heterostructure where color indicates the concentration of Ge atoms. Bottom: top-down view of a lateral cut through layer $l$ with cell discretization. Due to alloy disorder, different $p$ orbitals have different valley splittings.}
    \label{fig: heter_w_disorder}
\end{figure}
However, the analysis of Ref.~\cite{caporaletti_ProposedFiveElectronCharge_2025} assumed that the valley splitting due to the interface of the strained SiGe/Si/SiGe well is much greater than the $pO$ qubit splitting such that valley-orbit coupling (VOC) could be neglected. In this work, we remove this assumption by incorporating realistic alloy disorder and non-negligible VOC. Alloy disorder induces VOC because it creates an interface profile that varies laterally, so different orbital states effectively have different valley splittings. This is depicted in Fig.~\ref{fig: heter_w_disorder}, and in App.~\ref{App: VOC_theory} we derive a VOC strength $\sim1$--$10~\text{GHz}$ using the discretized effective-mass theory of alloy disorder from Ref.~\cite{losert_PracticalStrategiesEnhancing_2023}.

\begin{figure}[t]
    \centering
    \includegraphics[width = .9\columnwidth]{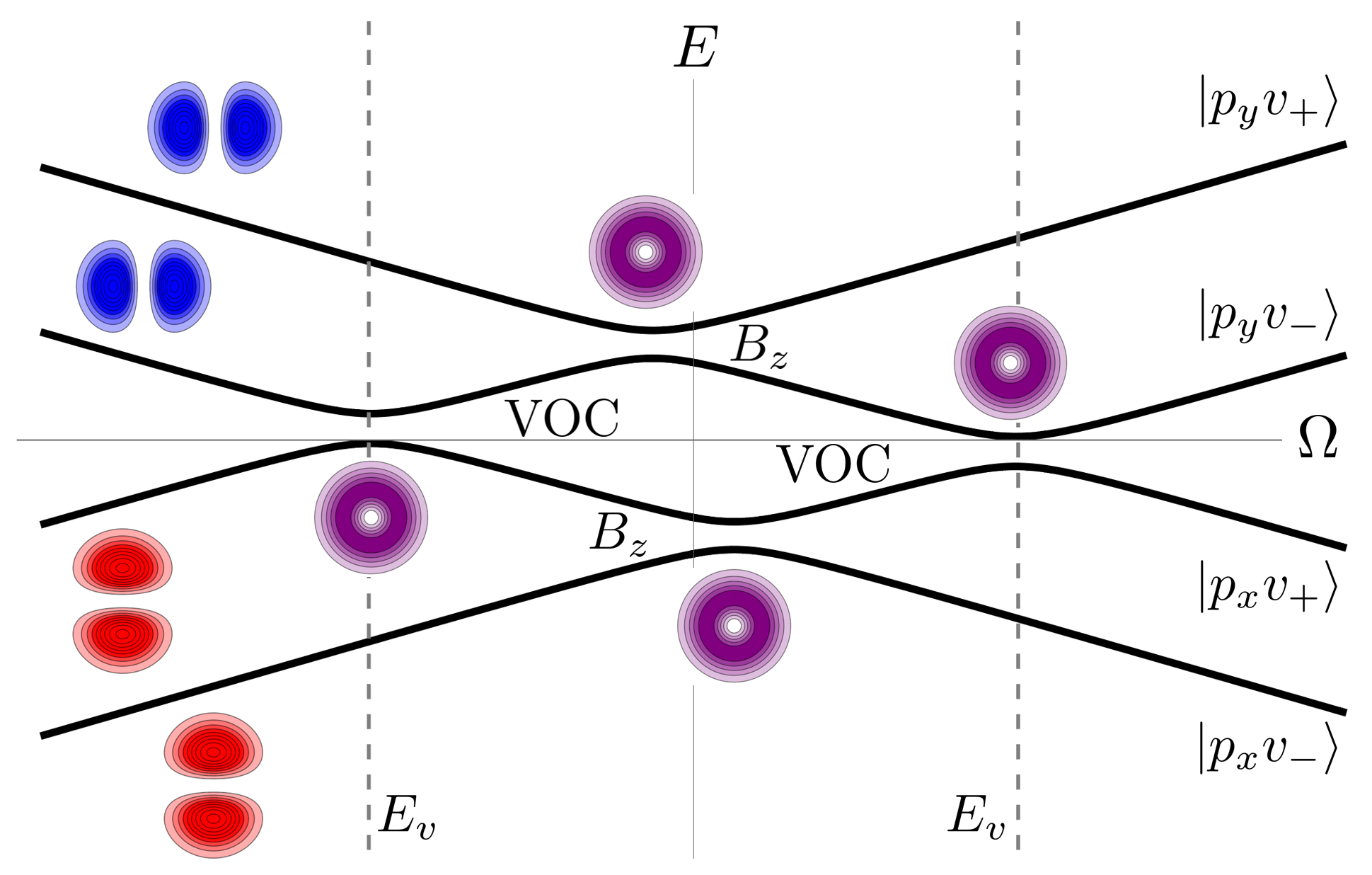}
    \caption{Schematic of the low-lying spectrum as a function of dot anisotropy for a fixed orientation, displaying four avoided crossings that function as sweet spots against charge noise. The 2D charge distributions of the  eigenstates are shown in red and blue on the left side, while the valley-orbit labels are shown on the right side. The charge distributions of any eigenstate at a sweet spot are shown in purple, with the level coupling source labeled.}
    \label{fig: pv_manifold}
\end{figure}

In the presence of VOC, we find that four sweet spots exist, defined as points where, in addition to the subspace's protection against uniform electric field fluctuations, an energetically isolated pair of states is first-order insensitive to fluctuations in the gradient of the electric field, or in other words, in the dot anisotropy and orientation. Two of the sweet spots are caused by an out-of-plane magnetic field mixing the $p_x$ and $p_y$ orbitals of the same valley (the previously considered $pO$ qubit manifold), while the other two are caused by VOC mixing the $p_x$ and $p_y$ orbitals of different valleys, as depicted in Fig.~\ref{fig: pv_manifold}. We dub an encoding into the two levels at this latter type of avoided crossing a $p$-orbital--valley, or $pOv$, qubit. 

Physically, each sweet spot is robust to charge noise because the eigenstates there have identical two-dimensional charge distributions. We show that operating at a sweet spot improves the dephasing time by roughly two orders of magnitude, $T_2^*\approx40~\text{ns} \rightarrow 10~\mu\text{s}$. Furthermore, we find that the $pOv$ qubit has a fully tunable two-qubit Coulomb interaction, unlike the always-on interaction of the $pO$ qubit considered previously~\cite{caporaletti_ProposedFiveElectronCharge_2025}.

This paper is structured as follows. In Sec.~\ref{sec: pO_intro}, we present an expanded introduction to the $pO$ qubit and its operation, parameterized in terms of the quantum dot eccentricity and orientation. In Sec.~\ref{sec: swt_spt_in_poV_spec}, we introduce the VOC Hamiltonian and calculate the effective two-level Hamiltonian of the $pO$ qubit in Sec.~\ref{subsec: pO_qubit_w_VOC} and of the $pOv$ qubit in Sec.~\ref{subsec: pOv_qubit}, along with the location of sweet spots in parameter space for each. An analytical, phenomenological charge-noise model derived from an ensemble of quasistatic electric dipoles is applied to each qubit to estimate the dephasing times when parked at the sweet spot. Finally, in Sec.~\ref{sec: two qubit}, we consider the quadrupole-quadrupole Coulomb interaction between $pOv$ qubits, finding that it is diagonal in the two-qubit basis and can be turned on and off by adjusting the relative orientations of adjacent anisotropic dots.

\section{$pO$-qubit Hamiltonian without VOC}\label{sec: pO_intro}
Consider an electron of effective mass $m^*$ in a harmonic, isotropic, two-dimensional quantum dot with Hamiltonian
\begin{equation}\label{eq: pO_H0}
   H_0 = \frac{\hbar \omega_0}{2}\left(-\nabla^2 + x^2+y^2\right),
\end{equation}
where $\hbar\omega_0$ is the confinement energy and $(x,y)$ are dimensionless coordinates scaled by the effective length $l_0=\sqrt{\frac{\hbar}{m^*\omega_0}}$. The eigenstates $\ket{n,m}$ are labeled by the shell index $n$ and the angular momentum projection $m\hbar$, with energies $\hbar\omega_0 (n+1)$ and shell degeneracy $n+1$.

To form the $pO$ qubit in a Si quantum dot with valley splitting smaller than the confinement energy, one fills the lowest ($s$) shell with four electrons (accounting for the spin and valley DOF) and places the fifth into the doubly degenerate ($p$) shell. The $pO$-qubit is encoded in this degenerate manifold, spanned by orbital angular momentum eigenstates $\ket{p_\pm}$ satisfying $L\ket{p_\pm}=\pm\hbar\ket{p_\pm}$. Because $\ket{p_\pm}$ differ by two units of angular momentum, the qubit has no dipole moment---the defining feature of the $pO$-qubit and the origin of its weak coupling to charge noise~\cite{caporaletti_ProposedFiveElectronCharge_2025}.

The shell-filling picture implicitly assumes that electron-electron interactions are negligible compared to the confinement energy, which is not the case for experimentally relevant dot sizes of $10$--$100~\text{nm}$~\cite{zajac_ScalableGateArchitecture_2016a}. Nevertheless, the noninteracting single-particle orbitals remain a useful basis because electron-electron interactions preserve the rotational symmetry of the dot, guaranteeing a doubly degenerate $p$-like manifold with no dipole moment in the interacting spectrum. Even in the strongly interacting Wigner-molecule limit~\cite{perezfadon_InteractioninducedSymmetryBreaking_2025}, a $pO$-qubit can still be defined; the interacting $p$-like Wigner molecule states may differ in size and radial structure from their noninteracting counterparts, but they retain the same angular momentum. In the remainder of this paper, when matrix elements must be computed, we use non-interacting orbitals to roughly estimate the values, recognizing that exact values require a much more demanding full configuration interaction computation.

One subtlety remains: with strong interactions, level crossings between states of different orbital angular momentum can shift the $p$-like manifold out of the ground state, as shown for a three-electron dot in Ref.~\cite{mikhailov_QuantumdotLithiumZero_2002}. We assume here that the interacting $p$-like manifold is either the ground manifold or a sufficiently long-lived metastable excited manifold. Long lifetimes in the latter case are plausible because relaxation out of the $p$-like manifold may require crossing spin sectors, and spin relaxation in Si is slow. A full configuration interaction computation of the five-electron Si dot including valleys is left for future work.

In light of this discussion, we use the single particle $p$-orbitals
\begin{equation}\label{eq: pO_basis}
    \ket{p_\pm}\equiv\ket{1,\pm1}
\end{equation}
to model $pO$-qubit manifold. One can perform arbitrary unitaries in this subspace by deforming the quantum dot along different axes in its plane, lifting the degeneracy. The electrical control Hamiltonian is
\begin{equation}\label{eq: H_eps}
    H_\epsilon=-\frac{\hbar \omega_0 }{2}\epsilon^2\left(x\sin\alpha - y \cos\alpha \right)^2,
\end{equation}
where $0 \leq \epsilon < 1$ is the eccentricity of the resulting elliptical confinement, while $-\frac{\pi}{2}<\alpha\leq\frac{\pi}{2}$ is the angle between the semi-minor axis and the $x$-axis. An out-of-plane magnetic field, $\textbf{B} = B_z \hat{\textbf{z}}$, adds a static control term to the Hamiltonian,
\begin{equation}
    H_B= -i \omega_0 l_0^2 e \textbf{A}\cdot \vec{\nabla}+ \frac{e^2 l_0^2}{2 m^*}\textbf{A}\cdot\textbf{A},
\end{equation}
where $e$ is the charge of the electron and we choose the symmetric gauge for the magnetic vector potential, $\mathbf{A}=\frac{B_z}{2}(-y\hat{\textbf{x}}+x\hat{\textbf{y}})$.

\begin{figure}[t]
    \centering
    \includegraphics[width=\columnwidth]{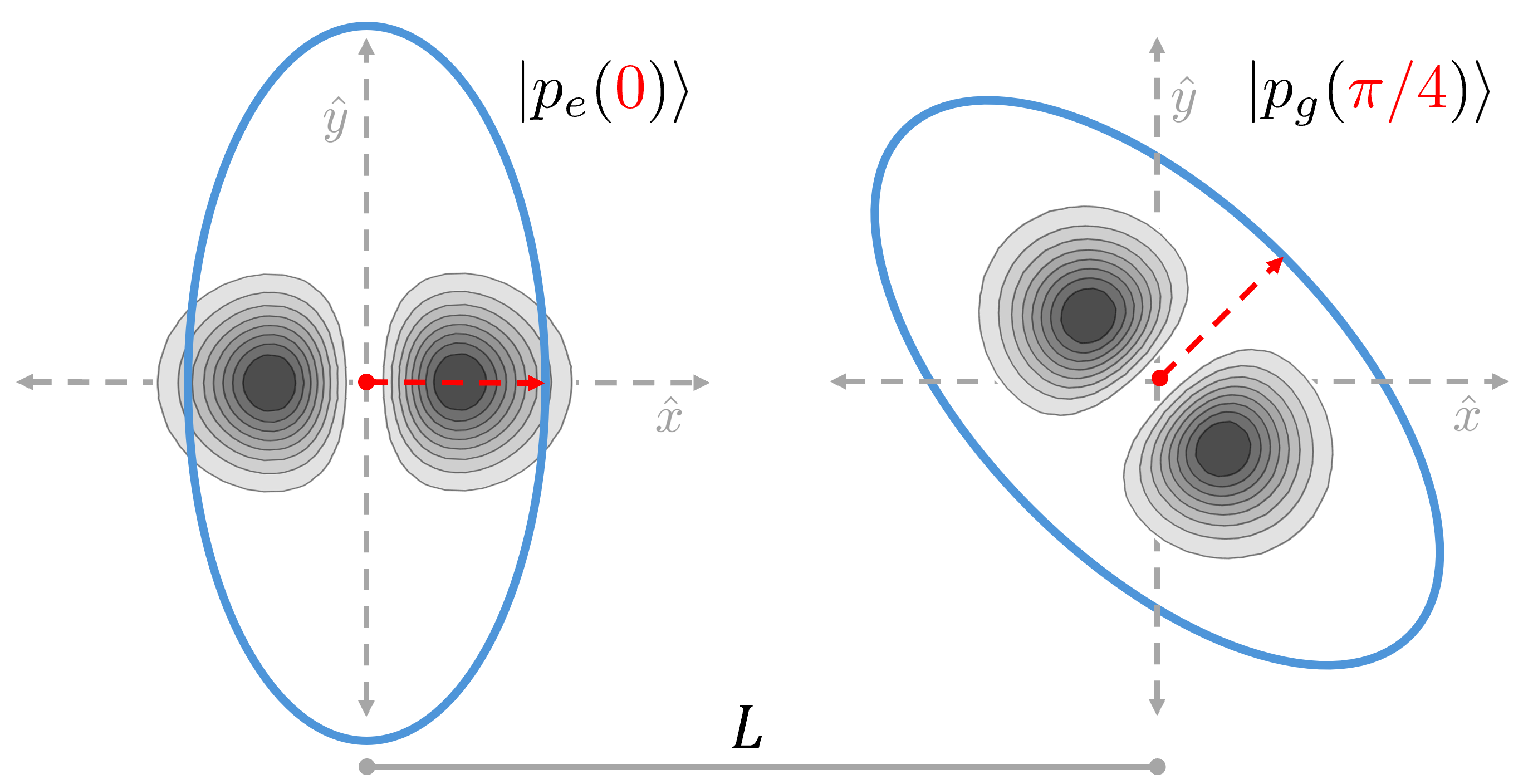}
    \caption{Adjacent quantum dots where the excited (ground) $p$-orbital charge distribution is depicted in the left (right) dot. The orientation of each dots semi-minor axis is shown via the red dashed arrow.}
    \label{fig: pOv_two_qubit_cartoon}
\end{figure}

We now project $H_0+H_\epsilon+H_B$ into the basis of Eq.~\eqref{eq: pO_basis}, yielding
\begin{equation}\label{eq: pO_H}
    H_{pO} \equiv H_0+H_\epsilon + H_B = \frac{\Omega}{2}\left(\cos(2\alpha)\sigma_x+\sin(2\alpha)\sigma_y\right) + \frac{E_z}{2}\sigma_z,
\end{equation}
where $\Omega=\frac{\hbar\omega_0}{2}\epsilon^2$, $E_z = 2 e B_z \hbar/2 m^*$, and $\sigma_{\ast}$ are the $\mathfrak{su}(2)$ generators for the $pO$ subspace. When $B_z=0$, the eigenstates of Eq.~\eqref{eq: pO_H} are
\begin{equation}\label{eq: beta_p_basis}
    \ket{p_{e/g}(\alpha)}\equiv\frac{1}{\sqrt{2}}\left(\ket{p_+} \pm e^{i 2\alpha}\ket{p_-}\right),
\end{equation}
where $\ket{p_e},\ket{p_g}$ are the familiar $p_x$ and $p_y$ orbitals, but spatially rotated by $\alpha$ relative to the $x$-axis, as shown in Fig.~\ref{fig: pOv_two_qubit_cartoon}.

\section{Effects of VOC on the $p$-orbital-valley manifold}\label{sec: swt_spt_in_poV_spec}
\begin{figure*}[t]
    \centering
    \includegraphics[width=1\textwidth]{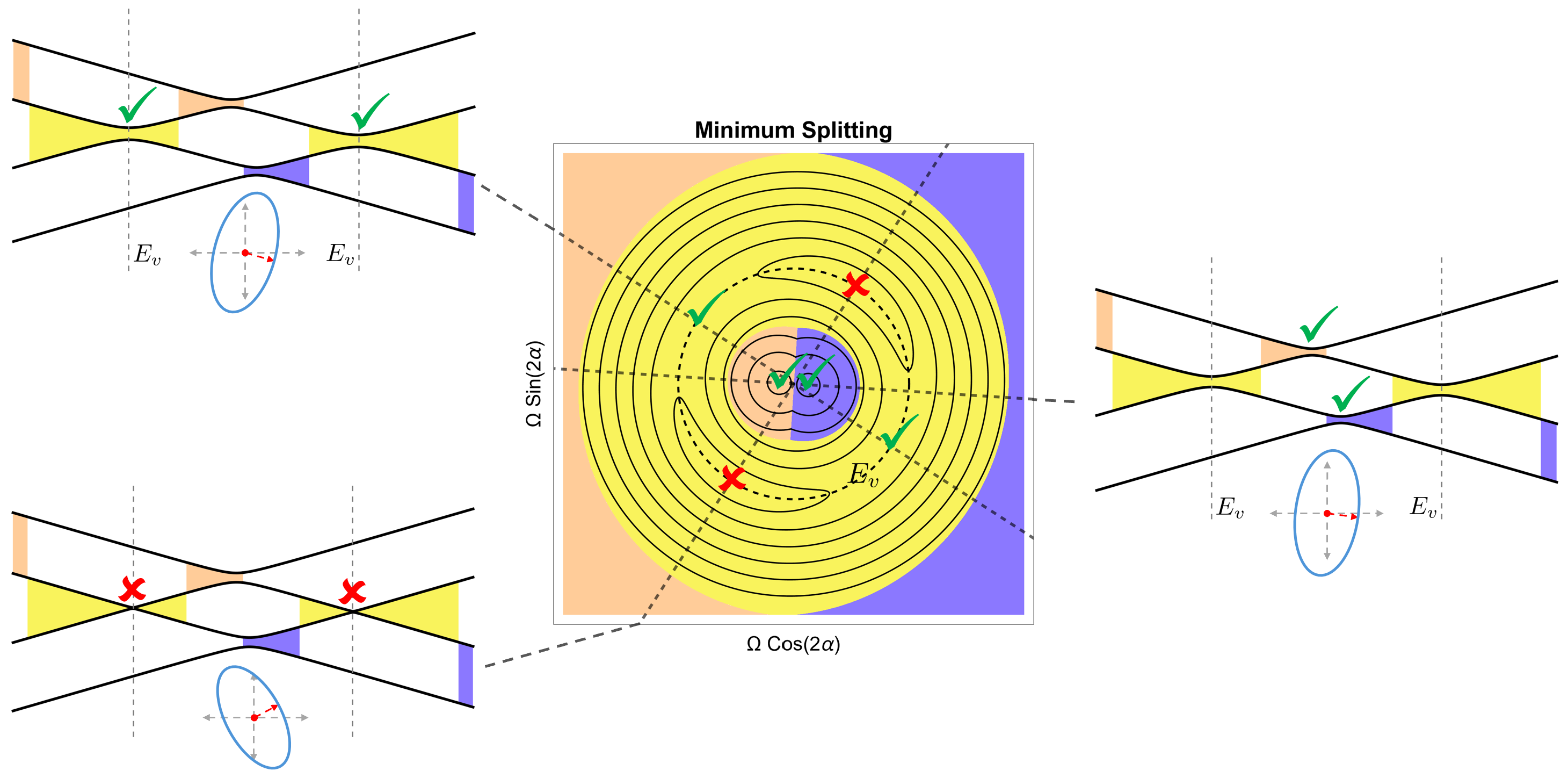}
    \caption{The center panel is a polar contour plot of the minimum of the 3 level splittings between the 4 eigenenergies of Eq.~\eqref{eq: H_main} vs.~dot anisotropy and orientation. The background color denotes which pair gives the minimal splitting, corresponding to the colors in the line cuts of the spectrum shown in the side panels. The line cuts are taken along fixed dot orientations marked by dashed lines in the center panel. Sweet spots are marked with green check marks and level crossings  are marked with red crosses. Vertical dashed lines in the spectrum cuts and the circular dashed line in the contour plot mark where $\Omega=E_v$. The dot orientations corresponding to the line cuts are shown beneath each spectrum with exaggerated anisotropy for clarity.}
    \label{fig: min_split_contour}
\end{figure*}
We now generalize the system model presented thus far to include valleys and VOC originating from alloy disorder at the well interface.
Electrons in Si possess six degenerate valley states at $\pm k_0 \hat{n}$ in momentum space, where $\hat{n}$ is one of $\{\hat{x}, \hat{y}, \hat{z}\}$, $k_0 = 0.82~(2\pi/a_0)$, and $a_0 = 0.543~\text{nm}$ is the width of the conventional cubic unit cell of Si. When the Si is in a vertically stacked Si/SiGe heterostructure, lateral strain due to the lattice mismatch between Si and SiGe lifts the six-fold degeneracy, leaving the two low-lying states at $\pm k_0 \hat{z}$. The heterostructure also provides a vertical confinement potential $U(z)$ originating from a conduction-band offset of the SiGe compared to the Si well. The two low-lying valleys are coupled through the atomic-scale features of $U(z)$~\cite{boykin_Valleysplittingstrained_2004}, producing a valley splitting $E_v$.

In the presence of disorder such that the confinement potential depends on the lateral position, $U(z) \rightarrow U(x,y,z)$, the valley splitting effectively depends on the electron's lateral orbital state, i.e., there is valley-orbit coupling (VOC). We estimate in App.~\ref{App: VOC_theory} that VOC has a strength of $\sim 1$--$10~\text{GHz}$ for realistic devices using the discretized, effective-mass theory in Ref.~\cite{losert_PracticalStrategiesEnhancing_2023}.

Incorporating valleys and VOC, the system Hamiltonian becomes
\begin{equation}\label{eq: H_main}
\begin{aligned}
    H(\Omega, \alpha) &= H_{pO} + H_v + H_\text{voc}\\
    &= \frac{\Omega}{2}\left(\cos\left(2\alpha\right)\sigma_x + \sin\left(2\alpha\right)\sigma_y \right) + \frac{E_z}{2} \sigma_z
    \\
    & \quad +\frac{E_v}{2} \left(\cos\theta_v \tau_x + \sin\theta_v \tau_y \right)
    \\
    & \quad +\eta_{xx} \sigma_x \tau_x + \eta_{xy} \sigma_x \tau_y + \eta_{yx} \sigma_y \tau_x + \eta_{yy} \sigma_y \tau_y,
\end{aligned} 
\end{equation}
where $\tau_\ast$ are the $\mathfrak{su}(2)$ generators for the valley subspace, $\theta_v$ is the valley phase, the form of $H_\text{voc}$ is derived in App.~\ref{App: VOC_theory}, but here it suffices to note that this is a completely generic coupling except that it contains no $\sigma_z$ terms. This is a result of the time-reversal-symmetric choice of basis together with the fact that $H_\text{voc}$ itself preserves time-reversal symmetry. Furthermore, the VOC coefficients $\eta_{**}$ are uncorrelated, Gaussian random variables with standard deviation $\sigma_\text{voc}$. Intuitively, the stochasticity is a direct consequence of the random nature of alloy disorder. However, for a given dot, provided it is not deformed so much during operations that the wavefunction samples different spatial regions at different times, the coefficients are constant. A more detailed discussion of these coefficients can be found in App.~\ref{App: VOC_theory}.

We will focus on disorder realizations such that the VOC is a small perturbation to $H_{pO}+H_v$,
\begin{equation}\label{eq: VOC_less_Ev}
    \overline{|H_\text{voc}|}\equiv\overline{E_\text{voc}}\lesssim \overline{E_v}/10,
\end{equation}
whose effect is mainly to open up avoided crossings when $\Omega = E_v$.
Here $|A| \equiv \frac{1}{\sqrt{d}}\sqrt{\text{Tr}(A^\dagger A)}$, $d$ is the dimension of matrix $A$, and the overline indicates an average over possible disorder realizations.

\subsection{The $pO$-qubit with VOC}\label{subsec: pO_qubit_w_VOC}
For a nearly isotropic dot with $\Omega \simeq 0$, far away from the $p$-orbital--valley level crossing at $\Omega=E_v$, entanglement between the valley and orbital DOF is negligible. We will consider only the ground-valley manifold, spanned by the states
\begin{equation}\label{eq: pO_v-_states}
    \ket{p_\pm^-}\equiv \ket{p_\pm}\otimes\ket{v_-},
\end{equation}
with
\begin{equation}\label{eq: valley_states}
    \ket{v_\pm}\equiv \frac{1}{\sqrt{2}}\left(\ket{v_{+k_0}} \pm e^{i \theta_v}\ket{v_{-k_0}}\right)
\end{equation}
the valley eigenstates of $H_v$ in Eq.~\eqref{eq: H_main} in terms of the uncoupled $z$-valley states at $\pm k_0$ in momentum space, $\ket{v_{\pm k_0}}$.

Projecting Eq.~\eqref{eq: H_main} into the basis of Eq.~\eqref{eq: pO_v-_states} yields
\begin{multline}
    H_{pO}^-(\Omega, \alpha) = \left(\frac{\Omega}{2} \cos(2\alpha) - a \right) \sigma_x 
    \\
    + \left(\frac{\Omega}{2} \sin(2\alpha) - b \right) \sigma_y 
    + \frac{E_z}{2} \sigma_z,
    \label{eq: H_pO_sub}   
\end{multline}
where $\sigma_\ast$ are here reinterpreted as the $\mathfrak{su}(2)$ generators in the $\ket{p^-_\pm}$ basis and $a$ and $b$ are constants determined by the particular disorder profile of the dot,
\begin{equation}
\begin{aligned}
    &a=\eta_{xx}\cos\theta_v+\eta_{xy}\sin\theta_v,\\
    &b=\eta_{yx}\cos\theta_v+\eta_{yy}\sin\theta_v.
\end{aligned}
\end{equation}
In addition to being immune to spatially uniform electric field fluctuations at any anisotropy, the $pO$ qubit is also immune to fluctuations in the dot anisotropy if
\begin{equation}
    \nabla \left|H_{pO}^-\right|_{\Omega_s, \alpha_s}=0,
    \label{eq: swt_spt_eq_pO}
\end{equation}
where $\vec{\nabla}\equiv\hat{\Omega}\partial_\Omega +\frac{1}{2}\hat{\alpha}\partial_{\alpha}$. Such a sweet spot exists at
\begin{equation}\label{eq: swt_spt_sol_v-}
\begin{aligned}
    &\Omega_s = 2\sqrt{a^2+b^2},\\
    &\alpha_s = \frac{1}{2}\arctan\left(a,b\right)
\end{aligned}
\end{equation}
when $E_z>0$, mixing the ground and excited $p$-orbital levels where they would otherwise cross. The effect of the alloy disorder here is simply to produce an intrinsic offset of the dot anisotropy by amount $\Omega_s$ with orientation $\alpha_s + \pi/2$, as shown in Eq.~\eqref{eq: H_pO_sub}. Therefore, operating at $\left(\Omega_s,\alpha_s\right)$ negates this offset, and is equivalent to operating at $(0,0)$ in the absence of VOC.

In App.~\ref{subApp: B_z swt spt dephasing} we estimate a dephasing time $T_2^*\sim10~\mu\text{s}$ at this sweet spot when $B_z= 20~\text{mT}$ (i.e., $E_z \approx 3~\text{GHz}$), matching the magnitude of magnetic dephasing due to fluctuating nuclear spins, as estimated in Ref.~\cite{caporaletti_ProposedFiveElectronCharge_2025} for a $800$ ppm $^{29}\text{Si}$ well \cite{chan_AssessmentSiliconQuantum_2018}. Increasing the value of $B_z$ would increase the robustness against charge noise, but the overall dephasing time would remain on the same order of magnitude due to the nuclear noise.

\subsection{The $pOv$-qubit}\label{subsec: pOv_qubit}
Now we consider $\Omega \simeq E_v$, where VOC mixes the ground $p$-orbital in the excited valley and the excited $p$-orbital in the ground valley. We expect this manifold to be metastable because relaxation to the ground $p$-orbital in the ground valley requires either the $p$-orbital to relax ($T_1\sim1~\text{s}$~\cite{caporaletti_ProposedFiveElectronCharge_2025}) or the valley to relax ($T_1\sim 10~\text{ms}$~\cite{penthorn_DirectMeasurementElectron_2020a}). The spectrum plotted in Fig.~\ref{fig: pv_manifold} corresponds to a special hand-picked case of VOC parameters that illustrates our main idea clearly, but a more general case, including the dependence on the dot orientation, is shown in Fig.~\ref{fig: min_split_contour}. Line cuts at various orientations are plotted in the side panels, while the central panel is a polar contour plot of the smallest energy splitting, $\min_{i} \left|E_{i+1} - E_i\right|$, vs dot anisotropy and orientation, color coded by which pair is the minimum at that point.

Near $\Omega=E_v$, the two intermediate-energy levels are closest to each other and are energetically separated from the lowest and highest levels by the valley splitting. This $pOv$ subspace is spanned by the states
\begin{equation}\label{eq: pOv_states}
\begin{aligned}
    &\ket{q_{g}(\alpha)}\equiv\ket{p_{g}(\alpha)}\otimes\ket{v_+},
    \\
    &\ket{q_{e}(\alpha)}\equiv\ket{p_{e}(\alpha)}\otimes\ket{v_-}.
\end{aligned}
\end{equation}
Note that $\braket{q_e(\alpha)|q_g(\alpha')} = 0,~\forall\alpha,\alpha'$, so when $\alpha$ is a function of time, even though the basis becomes time-dependent, there is no danger of inducing transitions.

The effective Hamiltonian is obtained by projecting Eq.~\eqref{eq: H_main} into the $\ket{q_{g/e}}$ basis,
\begin{equation}\label{eq: H_pOv}
    \begin{aligned}
    H_{pOv}(\Omega,\alpha) &= \Delta(\alpha)\sigma_x+\frac{\Omega-E_v}{2}\sigma_z,
    \end{aligned}
\end{equation}
where $\Delta(\alpha)\equiv c \cos(2\alpha)+d \sin(2\alpha)$, $\sigma_\ast$ are here reinterpreted as the $\mathfrak{su}(2)$ generators in the $\ket{q_{g/e}}$ basis and 
\begin{equation}
\begin{aligned}
    &c=\eta_{yy}\cos\theta_v - \eta_{yx}\sin\theta_v,\\
    &d=\eta_{xx}\sin\theta_v - \eta_{xy}\cos\theta_v.
\end{aligned}
\end{equation}
Sweet spots where
\begin{equation}
    \nabla \left|H_{pOv}\right|_{\Omega_s,\alpha_s}=0
    \label{eq: swt_spt_eq_pOv}
\end{equation}
exist at
\begin{equation}\label{eq: swt_spt_sol}
\begin{aligned}
    &\Omega_s=E_v,\\
    &\alpha_s=\frac{1}{2}\left(\arctan\left(c,d\right)+ n\pi\right),
\end{aligned}
\end{equation}
with $n \in \{0,1\}$.

In App.~\ref{subApp: pOv_dephasing}, we analytically estimate the dephasing time $T_2^*\equiv\sqrt{2}\hbar/\text{std}[E]$~\cite{petersson_QuantumCoherenceOneElectron_2010,kawakami_ElectricalControlLonglived_2014} of the $pOv$-qubit to be
\begin{equation}\label{eq: T_2_pOv}
    T_{2}^*(\alpha) \approx \frac{\hbar\,|\Delta(\alpha)|}{\sigma^2},
\end{equation}
where $\sigma$ is the per-component standard deviation of the noise Hamiltonian and at the sweet spot, $T_2^{*}(\alpha_s)\approx\hbar \sqrt{c^2+d^2}/\sigma^2$. In App.~\ref{subApp: chrg_noise_mod} we estimate $\sigma/h\sim4~\text{MHz}$, and since the dephasing time depends on the local disorder seen by the quantum dot, we average over the VOC random variables $\eta_{\ast\ast}$ and the valley phase $\theta_v$ from App.~\ref{App: VOC_theory} to obtain $\overline{T_2^*}\approx 10~\mu\text{s}$, where we assume that all VOC generator coefficients are independent, Gaussian random variables with standard deviation $\sigma_\text{voc} = 1~\text{GHz}$ (corresponding to $\overline{E_\text{voc}} \approx 2~\text{GHz}$) and the valley phase is uniformly distributed. So, for our underlying assumption of Eq.~\eqref{eq: VOC_less_Ev} to remain valid, we must have $\overline{E_v}\gtrsim 20~\text{GHz} \sim 80~\mu\text{eV}$. This strength of VOC and charge noise will be used for the remainder of the paper.

Note that control of the dot orientation is not required for single-qubit operations, but it is required to tune a $pO$ or $pOv$ qubit to a sweet spot in the presence of VOC. Also, although complete single-qubit control could be achieved while remaining at the sweet spot by driving the eccentricity resonantly, similar to the resonantly driven double-dot charge qubit \cite{kim_MicrowavedrivenCoherentOperation_2015}, it may be more desirable to use only baseband control. As is clear from Eq.~\eqref{eq: H_pOv}, with baseband control, complete two-axis control requires deforming away from the sweet spot, where the dephasing time goes down to $\sim 40~\text{ns}$~\cite{caporaletti_ProposedFiveElectronCharge_2025}. Nonetheless, the qubit is protected from noise while performing $X$ rotations and therefore the average quality factor for a sequence of single-qubit gates generally increases. Furthermore, changing the dot anisotropy can produce a large range of $\Omega-E_v$ such that the $Z$ rotations away from the sweet spot can be performed as quickly as the bandwidth of the classical control electronics allows.

 % The expected splitting of the $pOv$ qubit with $\overline{E_\text{voc}}$ are both $E_z=2\overline{|H_q(c_s)|}\approx 3 \text{ GHz}$. This can be swept through diabatically using state-of-the-art arbitrary waveform generators (AWG), which have a sample rate $10 \text{ GHz}$ \cite{kim_QuantumControlProcess_2014} and can even be done using more common AWG with $1~\text{GHz}$ sample rate using the sub-ns control developed in the Supplement of Ref.~\cite{vanriggelen-doelman_CoherentSpinQubit_2024}.
 
\section{Two-qubit system}\label{sec: two qubit}
Two-qubit gates are essential for universal quantum computation. In the following, we derive the $pO$ and $pOv$ two-qubit Hamiltonians in the presence of alloy disorder, each originating from the Coulomb interaction between valence electrons. It is found that the interaction between two $pO$-qubits is always on while the interaction between two $pOv$-qubits is tuneable by modulating each dot's orientation. For either qubit encoding, the dots are separated by a center-to-center distance $L$ as sketched in Fig.~\ref{fig: pOv_two_qubit_cartoon}. The distance can be chosen such that the $p$-orbitals of each dot has little overlap, making kinetic exchange (which depends on both the spin and orbital states of the electrons) negligible to avoid leakage into the spin subspace. Given the negligible wavefunction overlap, a product basis is used for both qubit encodings.

\subsection{Two $pO$-qubits}\label{sec: two-pO-qubit}
To calculate the Coulomb interaction between two $pO$-qubits, we will use the basis states
\begin{equation}\label{eq: 2pO_qubit_basis}
    \ket{p_{m}p_{n}} \equiv \ket{p_{m}}\otimes\ket{p_{n}},
\end{equation}
where $m=\pm1$, and $\ket{p_{\pm}}$ are the left and right circularly polarized $p$-orbitals defined in Eq.~\eqref{eq: pO_basis} which are also the sweet-spot eigenstates of Eq.~\eqref{eq: H_pO_sub}, and the different centers of the two dots is implicit.

The Coulomb matrix elements are
\begin{equation}\label{eq: pO_2Q_els}
    C_{mn}^{m^\prime n^\prime}\equiv\frac{e^2}{4\pi \epsilon}\braket{p_{m} p_{n} |\frac{1}{|\mathbf{r_1}-\mathbf{r_2}|} |p_{m^\prime} p_{n^\prime}}.
\end{equation}
Performing a multipole expansion, we can separate different contributions to the Coulomb interaction. The monopole-monopole term produces an identity operator, which is not relevant for operation of the qubits, and we neglect it. There are no dipole moments in this subspace, so the next order term is a monopole-quadrupole interaction, in which the state-independent total charge of one dot couples to the quadrupole moment of the other. The resulting single-qubit $\sigma_x$ terms can be calibrated away by tuning the top-gate voltages to compensate for this inter-dot deformation, so for simplicity we neglect this effect as well. The important contribution is the quadrupole-quadrupole interaction, which entangles the two quadrupolar charge qubits,
\begin{equation}\label{eq: pO_multipole_exp}
    C_{mn}^{m^\prime n^\prime}\approx\frac{e^2}{4\pi \epsilon} \frac{1}{4}Q_{ij}^{mn}Q_{k\ell}^{m^\prime n^\prime} \partial_i \partial_j \partial_k\partial_\ell \frac{1}{|\vec{r}|} \Bigg|_{\vec{r} = L\hat{x}},
\end{equation}
where
\begin{equation}
    Q_{ij}^{mn}\equiv \braket{p_{m}| r_i r_j |p_{n}}.
\end{equation}
is the quadrupole moment tensor of the single-particle wavefunction with respect to the center of the dot. Performing the sums in Eq.~\eqref{eq: pO_multipole_exp} and writing the resulting matrix in terms of Pauli generators produces the two-qubit interaction Hamiltonian,
\begin{equation}\label{eq: pO_2q_int}
\begin{aligned}
    H_{2q}
    &=\frac{e^2l_0^4}{2^6\pi \epsilon L^5}\left(57\sigma_x^{(1)}\sigma_x^{(2)}-48\sigma_y^{(1)}\sigma_y^{(2)}\right)
    \\
    &\equiv \Omega_{xx} \sigma_x^{(1)}\sigma_x^{(2)} + \Omega_{yy} \sigma_y^{(1)}\sigma_y^{(2)}.
\end{aligned}
\end{equation}
For the device parameters in Table~\ref{Tab: device_params}, these interaction terms are on the order of 1 GHz.

To control the qubit pair, one can either change the interdot distance by charge shuttling or one can accept always-on coupling and use shaped control pulses on the anisotropy and dot orientation to perform single- and two-qubit gates. In Ref.~\cite{caporaletti_ProposedFiveElectronCharge_2025}, pulse shape optimization is used to find the universal gate-set $\{H^{(k)}, S^{(k)}, T^{(k)}, \text{bSWAP}\}$, where $H^{(k)}$ is the Hadamard gate on qubit $k$, and likewise $S^{(k)}$ and $T^{(k)}$ are phase gates of angle $\frac{\pi}{2}$ and $\frac{\pi}{4}$ respectively, while $\text{bSWAP} = \exp{\left[i\frac{\pi}{4}\left(\sigma_x^{(1)}\sigma_x^{(2)}-\sigma_y^{(1)}\sigma_y^{(2)}\right)\right]}$ \cite{wei_NativeTwoQubitGates_2024} is a perfectly entangling gate \cite{zhang_GeometricTheoryNonlocal_2003} and is equivalent to $\text{iSWAP} = \exp{\left[i \frac{\pi}{4} \left(\sigma_x^{(1)}\sigma_x^{(2)}+\sigma_y^{(1)}\sigma_y^{(2)}\right)\right]}$ up to local $\sigma_x$ rotations \cite{poletto_EntanglementTwoSuperconducting_2012}. Using the same phenomenological noise model discussed in App.~\ref{subApp: chrg_noise_mod}, Ref.~\cite{caporaletti_ProposedFiveElectronCharge_2025} estimates gate infidelities $\sim 10^{-4}$ without a magnetic field and $\sim10^{-5}$ with a uniform magnetic field such that $E_z/h=4.5~\text{GHz}$. The lower infidelity at non-zero magnetic field is due to the sweet spot physics discussed in Sec.~\ref{subsec: pO_qubit_w_VOC}.
 
\subsection{Two $pOv$-qubits}\label{sec: two-pOv-qubit}
For two $pOv$-qubits coupled via the Coulomb interaction, the appropriate basis states are
\begin{equation}\label{eq: 2_qubit_basis}
    \ket{q_{m}\left(\alpha_1\right)q_{n}\left(\alpha_2\right)} \equiv \ket{q_{m}\left(\alpha_1\right)}\otimes\ket{q_{n}\left(\alpha_2\right)},
\end{equation}
where $\ket{q_{m}(\alpha)}$ is defined in Eq.~\eqref{eq: pOv_states} and $\alpha_i$ is the orientation of dot $i$, and the different centers of the two dots is implicit.

It is a good approximation to take the Coulomb interaction to be diagonal in this basis because terms which couple different valley states have an oscillating phase, $e^{i2k_0z}$, in the integrand that average out underneath the slowly varying product of the Coulomb potential and the envelope function of the valley states.
The Coulomb diagonal matrix elements are
\begin{equation}\label{eq: ZZ_els}
    C_{mn}\equiv\frac{e^2}{4\pi \epsilon}\braket{q_{m} \left(\alpha_1\right) q_{n} \left(\alpha_2\right)|\frac{1}{|\mathbf{r_1}-\mathbf{r_2}|} |q_{m} \left(\alpha_1\right) q_{n} \left(\alpha_2\right)}.
\end{equation}

\begin{figure}[t]
    \centering
    \includegraphics[width=.9\columnwidth]{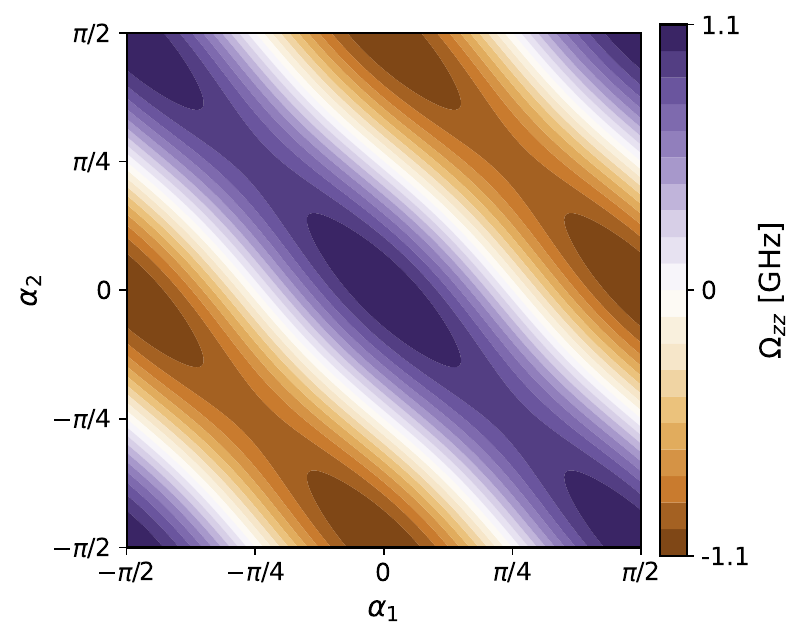}
    \caption{Quadrupole-quadrupole interaction strength as a function of each dots orientation using the parameters in Table~\ref{Tab: device_params}. The white stripes correspond to zero interaction.}
    \label{fig: chi}
\end{figure}

Again performing a multipole expansion of Eq.~\eqref{eq: ZZ_els} as we did in the previous section with Eq.~\eqref{eq: pO_2Q_els}, the entangling quadrupole-quadrupole interaction is
\begin{equation}\label{eq: multipole_exp}
    C_{mn}\approx\frac{e^2}{4\pi \epsilon} \frac{1}{4}Q_{ij}^{(m)}(\alpha_1)Q_{k\ell}^{(n)}(\alpha_2) \partial_i \partial_j \partial_k\partial_\ell \frac{1}{|\vec{r}|} \Bigg|_{\vec{r} = L\hat{x}},
\end{equation}
where
\begin{equation}
    Q_{ij}^{(m)}(\alpha)\equiv \braket{q_{m}(\alpha)| r_i r_j |q_{m}(\alpha)}.
\end{equation}
Summing over the indices in Eq.~\eqref{eq: multipole_exp} and writing the resulting matrix in terms of the Pauli generators produces
\begin{equation}\label{eq: H_2q_pOv}
    H_{2q} = \Omega_{zz}(\alpha_1,\alpha_2)\sigma_z^{(1)} \sigma_z^{(2)},    
\end{equation}
where 
\begin{equation}\label{eq: chi}
\begin{aligned}
    \Omega_{zz}(\alpha_1,\alpha_2)\equiv\frac{3 e^2 l_0^4}{2^7 \pi \epsilon L^5} &\bigg[3 \cos \big( 2\left(\alpha_1 -\alpha_2 \right) \big)\\
    &+ 35 \cos \big( 2\left( \alpha_1 + \alpha_2 \right) \big) \bigg].
\end{aligned}
\end{equation}
Therefore, by adjusting the orientations of the dots, the two-qubit interaction can be turned on and off. The coupling strength is shown in Fig.~\ref{fig: chi}. Although it would be useful if $\Omega_{zz}$ could be tuned while both qubits remained at their sweet spots, that is not the case. However, a compromise can often be made where both qubits remain close to their sweet spots while $\Omega_{zz}$ is set to a desired strength.

\begin{figure}[t]
    \centering
    \includegraphics[width=\columnwidth]{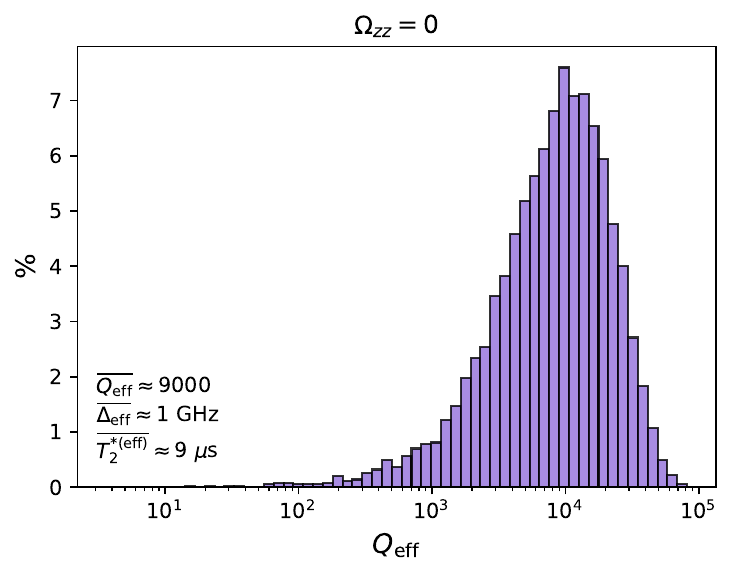}
    \caption{Histogram of $Q_\text{eff}$ when $\Omega_{zz}=0$ for $10^4$ disorder realizations.}
    \label{fig: Q_hist}
\end{figure}

Single-qubit operations are performed while $\Omega_{zz}=0$. As is clear from Eq.~\eqref{eq: chi} and Fig.~\ref{fig: chi}, there is a continuous set of orientations that satisfy this condition, so within this set we choose the one that maximizes
\begin{equation}\label{eq: Q_eff}
    Q_\text{eff} = \Delta_\text{eff}T_2^{*(\text{eff})}/h,
\end{equation}
where
\begin{equation}\label{eq: Delta_eff and T2_eff}
\begin{aligned}
    &\Delta_\text{eff}\equiv 2\left(\frac{1}{\Delta(\alpha_1)}+\frac{1}{\Delta(\alpha_2)}\right)^{-1},\\
    &T_2^{*(\text{eff})}\equiv 2\left(\frac{1}{T_2^{*}(\alpha_1)}+\frac{1}{T_2^{*}(\alpha_2)}\right)^{-1},    
\end{aligned}
\end{equation}
with $T_2^{*}$ and $\Delta$ defined in Eq.~\eqref{eq: T_2_pOv} and Eq.~\eqref{eq: H_pOv} respectively. A histogram of the optimized $Q_\text{eff}$ for $10^4$ disorder realizations is shown in Fig.~\ref{fig: Q_hist}. Low quality factors originate from disorder realizations with small values $\max[\Delta]=\sqrt{c^2+d^2}$, resulting in slow gate times and poor dephasing.

The native entangling gate is
\begin{equation}\label{eq: pulse_seg}
U= \exp\left[\frac{-i t}{\hbar}\left(\Delta_{1}\sigma_x^{(1)}+ \Delta_{2}\sigma_x^{(2)} + \Omega_{zz}\,\sigma_z^{(1)}\sigma_z^{(2)}\right)\right],
\end{equation}
where the anisotropy of each qubit has been set to its valley splitting such that the single-qubit $\sigma_z$ terms from Eq.~\eqref{eq: H_pOv} are zero and $\Delta_{i}$ and $\Omega_{zz}$ are functions of the two orientations $\alpha_i$. 

We wish to produce a gate $R_\text{ZZ}(\pi/2) \equiv e^{-i(\pi/4)\sigma_z^{(1)}\sigma_z^{(2)}}$, locally equivalent to a CZ gate. This is achieved by rotating each dot orientation to one of the zeroes of its $\Delta$, denoted as $\beta$ (marked by crosses in Fig.~\ref{fig: min_split_contour}), for a time 
\begin{equation}\label{eq: N=1_time}
    t=\frac{h}{8\Omega_{zz}(\beta_1,\beta_2)}.
\end{equation}

\begin{figure}[t]
    \centering
    \includegraphics[width=\columnwidth]{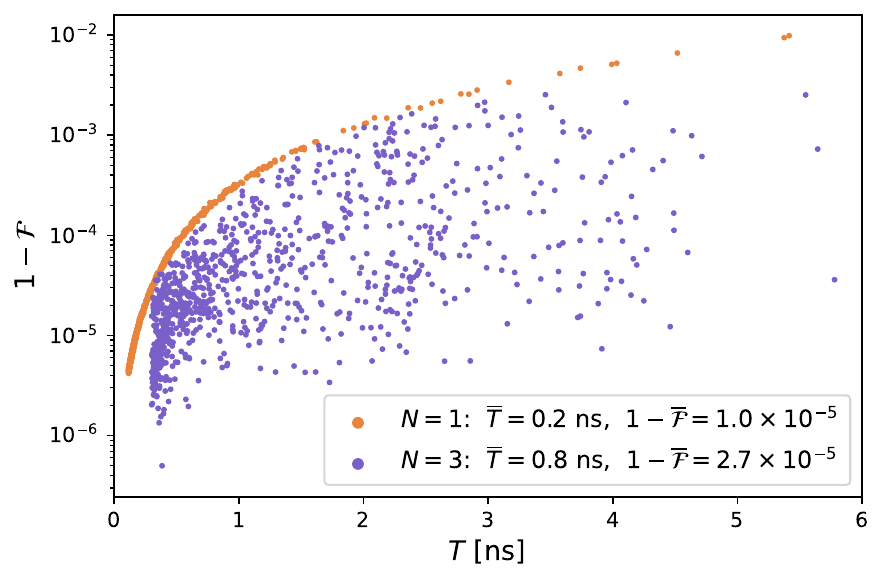}
    \caption{Scatter plot of infidelity $1-\mathcal{F}$ for $N=1,3$ applications of Eq.~\eqref{eq: pulse_seg} with optimized parameters for $10^3$ disorder realizations. About $2\%$ of $N=1$ cases lie outside the plotted domain and have $1-\mathcal{F}\geq10^{-2}$. The median total gate time $\overline{T}$ and median infidelity $1-\overline{\mathcal{F}}$ are reported in the legend.}
    \label{fig: inf_plot}
\end{figure}

The gate infidelity due to quasistatic charge noise is
\begin{equation}\label{eq: noise_inf}
    1-\mathcal{F}=1-\frac{1}{16}\mathbb{E}\left[\left|\mathrm{Tr}\left(R_\text{ZZ}^\dagger\left(\frac{\pi}{2}\right)\tilde{U}\right)\right|^2\right],
\end{equation}
where $\tilde{U}$ includes perturbations to the dot orientations $\alpha$, anisotropies $\Omega$, and interdot distance $L$, and $\mathbb{E}$ denotes the average over charge noise configurations. See App.~\ref{subApp: two_pOV_pulse_inf} for details about generating the charge noise potential for a given configuration and translating it to perturbed dot parameters. We numerically average over $10^3$ charge noise realizations and show the infidelities for $10^3$ alloy disorder realizations in Fig.~\ref{fig: inf_plot}. Note that $\sim2\%$ of the alloy disorder realizations lie at longer times than shown on the plot and have infidelity above $10^{-2}$, due to their values of $\beta$ happening to lie close to the $\Omega_{zz}=0$ curve in Fig.~\ref{fig: chi}. This is unlikely to appear in a small device, and is merely an annoyance in a larger device that has a geometry that allows certain links to be avoided. Alternatively, though, we show below that one can eliminate this problem by instead using a three-segment composite pulse. 

For a pulse sequence composed of three applications of Eq.~\eqref{eq: pulse_seg}, each with different values of $\alpha$, one no longer needs to operate at the roots of $\Delta$ because the single-qubit $X$ terms of one segment can be canceled by those of a different segment. This allows $\Omega_{zz}$ to be operated near its maximum, resulting in generally faster pulse sequences with low infidelity. We minimize the total pulse time $T\equiv\sum_{k=1}^3 t_k$, subject to the segment bounds $100~\text{ps} \leq t_k \leq 2~\text{ns}$, while enforcing that infidelity in the absence of noise be $\leq 10^{-13}$ using the optimization method SLSQP. The pulse infidelity due to charge noise is shown in Fig.~\ref{fig: inf_plot} for $10^3$ alloy disorder realizations, all of which have infidelity below $10^{-2}$.

The three-segment pulse always outperforms any single-segment pulse of the same time (Fig.~\ref{fig: inf_plot}), since retaining a nonzero $\Delta$ provides the sweet-spot-like protection against charge noise as discussed for the single-qubit case in Sec.~\ref{sec: swt_spt_in_poV_spec}. However, the single-segment pulse achieves a lower overall median infidelity simply because the $100~\text{ps}$ floor on each segment means the three-segment sequence can never run in less than $300~\text{ps}$, whereas for many disorder realizations the single-segment pulse completes well below this time. These faster gates accumulate less dephasing, producing a better median infidelity.

In practice, the optimal pulse parameters for a specific alloy-disorder realization can be found in two ways. The avoided crossing strengths, $\Delta_i$, can first be experimentally mapped as a function of dot orientations, and then the optimal orientations calculated as above. Alternatively, one could skip the mapping and directly carry out an in situ optimization of infidelity over the pairs of orientations experimentally. This closed-loop approach would optimize against the noise the device actually experiences. This would be a one-time pairwise calibration of the device.

\section{Conclusion}\label{sec: conclusion}
This work has generalized the $pO$ qubit theory developed in Ref.~\cite{caporaletti_ProposedFiveElectronCharge_2025} to include VOC. Within this extended formalism we defined the $pOv$ qubit, an encoding in the two-level manifold spanned by the ground $p$-orbital in the excited valley and the excited $p$-orbital in the ground valley. The $pOv$ qubit exhibits sweet spots where the vertically-integrated charge distributions of the instantaneous eigenstates are identical, giving rise to further charge noise resilience. Both the strength and location in control space of the sweet spots depend on the local alloy disorder of the dot. Therefore, the sweet spot dephasing time will vary from dot to dot. We estimate a site-averaged qubit dephasing time of $T_2^*\approx 10~\mu\text{s}$ when parked at the sweet spot, paired with GHz Rabi rates.

We have also shown an alternative route to obtaining $T_2^*\approx 10~\mu\text{s}$ using the previously defined $pO$ qubit with an out-of-plane magnetic field, $B_z$. Applying $B_z$ makes the instantaneous eigenstates of the system the left- and right-circularly polarized $p$-orbital states, which have identical charge distributions, leading again to a deterministic sweet spot with enhanced resilience to charge noise. It is shown that VOC merely offsets the anisotropy and orientation defining the sweet spot location.

Additionally, we have shown that neighboring $pOv$ qubits couple through the quadrupole-quadrupole Coulomb interaction, similar to the $pO$ qubits. This coupling is diagonal in the two-qubit basis, realizing a native Ising ($\sigma_z^{(1)}\sigma_z^{(2)}$) interaction whose strength is set by the relative orientation of the two anisotropic dots. It is therefore fully electrically tunable from zero to $\sim1~\text{GHz}$, switching off during single-qubit operations and on to generate entanglement, in contrast to the always-on interaction of the $pO$ qubit. Synthesizing an $R_{ZZ}(\pi/2)$ gate from baseband square pulses under the same TLF charge-noise model used to estimate $T_2^*$, we obtained sub-nanosecond gate times with median charge-noise infidelities $\sim10^{-5}$ across the sampled disorder realizations.

Due to the stochastic nature of alloy disorder at the interface, operating an array of either type of qubit requires characterization of the valley splitting and of the relevant avoided crossing gap as a function of dot anisotropy and orientation. Valley splitting measurement is a standard experiment by now and measuring the avoided crossings could be done in a similar way. It also requires gate infrastructure capable of independently deforming each dot's orientation so that the optimal operating points can be reached. However, these scalability challenges may be offset by the ability to tune two-$pOv$-qubit interactions via modulating the relative orientation of adjacent dots. Further consideration of operating scalable arrays of $pO$/$pOv$ qubits will be the subject of future work.

\begin{acknowledgements}
    The authors thank Dr. Yasuo Oda for helpful discussions while developing this work and acknowledge support from the Army Research Office (ARO) under Grant No. W911NF-23-1-0115 
\end{acknowledgements}

\section{Appendix}\label{sec: appendix}
All numerical simulations in this paper use the following device parameters given in Table~\ref{Tab: device_params}.
\begin{table}[h]
\centering
\begin{tabular*}{\columnwidth}{l c c}
 \hline
 Parameter & Symbol & Value \\
 \hline 
  Characteristic Dot Length& $l_0$ & $10$ nm \\
  Inter-dot Distance & $L$ & $63$ nm\\
  Vertical Electric Field& $F$ & $7.5~\text{mV}/\text{nm}$~\cite{paqueletwuetz_AtomicFluctuationsLifting_2022}\\
  Interface Width& $\lambda_{\text{int}}$ & $6$ ML~\cite{paqueletwuetz_AtomicFluctuationsLifting_2022} \\
  Well Width& $W\equiv z_t-z_b$& $ 120$ ML~\cite{paqueletwuetz_AtomicFluctuationsLifting_2022}\\
  Substrate (Well) Si & $X_s (X_w)$ & $0.665(1)$~\cite{paqueletwuetz_AtomicFluctuationsLifting_2022}\\
  Relative Permittivity of Si& $\epsilon_r$& $11.6$~\cite{szePhysicsSemiconductorDevices2021}\\
  Longitudinal Effective Mass&$m^*_l$&$0.98 m_e$~\cite{szePhysicsSemiconductorDevices2021}\\
  Transverse Effective Mass&$m^*_t$&$0.19 m_e$~\cite{szePhysicsSemiconductorDevices2021}\\
  \hline
  Dot Distance to Electrodes& $D$ & $100$ nm \\
  TLF Dipole Length& $d$ & $0.1$ nm~\cite{reinisch_MicroscopicDescriptionLowtemperature_2006} \\
  TLF Dipole Density& $\rho$ & $10^{-3}~\text{nm}^{-2}$~\cite{zimmermann_Thermalrelaxationlowenergy_1981} \\
  \hline
\end{tabular*}
\caption{Parameter set which characterize the Si/SiGe device used for all calculations. The upper portion of the table defines fabrication and operating parameters, while the lower portion defines two-level fluctuator (TLF) defect parameters.}
\label{Tab: device_params}
\end{table}
\subsection{Valley-Orbit Coupling Theory}\label{App: VOC_theory}
In this section, we quantify disorder-induced VOC within the $p$-orbital subspace using the discretized, effective-mass theory developed in Ref.~\cite{losert_PracticalStrategiesEnhancing_2023}. In strained SiGe/Si/SiGe heterostructures, interface steps and alloy concentration fluctuations are the primary sources of disorder~\cite{kharche_Valleysplittingstrained_2007}. However, when the Si/SiGe interface width is $\lambda_\text{int} \geq 0.5~\text{nm}$, which appears to be the case in practice~\cite{degliesposti_LowDisorderHigh_2024}, interface steps produce negligible VOC~\cite{losert_PracticalStrategiesEnhancing_2023}. Therefore, we exclusively consider alloy disorder in our theoretical treatment.

First, consider discretizing the SiGe/Si/SiGe heterostructure into cells $C_{j,k,l}$ of dimension $(\Delta x,\Delta y, \Delta z)$ with centers $\textbf{r}_{j,k,l} \equiv (x_j,y_k,z_l)$.
 Ref.~\cite{losert_PracticalStrategiesEnhancing_2023} posits that the confinement potential energy in each cell is an independent, stochastic variable given by
\begin{equation}\label{eq: U_cell}
    U_{j,k,l} \equiv \Delta E_c\frac{X_{j,k,l} - X_s}{X_w-X_s} 
\end{equation}
where $\Delta E_c$ is conduction band offset,
\begin{equation}\label{eq: Si_concen}
    X_{j,k,l}\equiv\frac{\text{Binom}(N_\text{c},\overline{X_l})}{N_\text{c}}
\end{equation}
is a binomial random variable representing the concentration of Si in $C_{j,k,l}$, $X_{w(s)}$ is the mean Si concentration at the center of the Si well (deep in the SiGe substrate),
\begin{equation}
    \overline{X_l} = \left(X_w + \frac{X_s-X_w}{1+e^{\frac{z_l-z_t}{\tau}}} + \frac{X_s-X_w}{1+e^\frac{z_b-z_l}{\tau}}\right)
\end{equation}
is the mean Si concentration profile of the well, formed by two sigmoidal interfaces with centers $z_t,z_b$ and width $\lambda_\text{int}\equiv4\tau$, and $N_\text{c} = \frac{8}{a_0^3}V_c$ is the number of atoms in a cell where $\frac{8}{a_0^3}$ is the density of atoms in the conventional unit cell of Si and $V_c=\Delta x\Delta y\Delta z$. Although the stochastic, discretized confinement energy in Eq.~\eqref{eq: U_cell} has a mean value
\begin{equation}\label{eq: U_det}
    U_l\equiv\overline{U_{j,k,l}} = \Delta E_c\frac{\overline{X_l}-X_s}{X_w-X_s}
\end{equation}
that is independent of the lateral coordinates, its finite variance
\begin{equation}
\begin{aligned}
    \overline{\delta U_{j,k,l}^2} &\equiv \overline{(U_{j,k,l}-\overline{U_{j,k,l}})^2}\\
    &= \frac{\Delta E_c^2}{(X_w-X_s)^2 N_\text{c}} \overline{X_l}(1-\overline{X_l})
\end{aligned}
\end{equation}
causes the confinement energy to vary randomly from cell to cell. Therefore, the mean confinement energy $U_l$ acts only on the valley space to produce deterministic valley splitting $E_{v0}$, while the fluctuation away from the mean $\delta U_{j,k,l}$ is the physical source of VOC. We now project $\delta U_{j,k,l}$ into the $p$-orbital--valley subspace:
\begin{equation}\label{eq: U_disorder}
\begin{aligned}
    \delta U_{m,n}^{m^\prime,n^\prime}
    = V_c \braket{\Psi_{m,n}|\textbf{r}_{j,k,l}}\delta U_{j,k,l}\braket{\textbf{r}_{j,k,l}|\Psi_{m^\prime,n^\prime}},    
\end{aligned}
\end{equation}
where repeated indices are summed over and the time-reversal-symmetric $p$-orbital--valley basis states are given by
\begin{equation}\label{eq: raw_p_val_basis}
\begin{aligned}
    &\ket{\Psi_{m,n}}= \ket{p_m} \otimes \ket{v_n}\equiv\ket{p_mv_{n k_0}},\\
    &\braket{\textbf{r}|v_{nk_0}} = \phi_0(z)e^{n i k_0 z},
\end{aligned}
\end{equation}
where $\ket{p_m}$ comes from Eq.~\eqref{eq: pO_basis}, $\ket{v_{nk_0}}$ represents the low-lying valley states, $m,n\in \pm$, and $\phi_0(z)$ is the ground state of the deterministic vertical confinement. Normalization is enforced via $V_\text{c}\sum_{j,k,l}|\braket{\textbf{r}_{j,k,l}|\Psi_{m,n}}|^2 = 1$, and $\ket{\textbf{r}_{j,k,l}}$ represents a discrete orthonormal position basis ket with inner product $\braket{\textbf{r}_{j,k,l}|\textbf{r}_{j^\prime,k^\prime,l^\prime}} = \frac{1}{V_c}\delta_{j,j^\prime} \delta_{k,k^\prime} \delta_{l,l^\prime}$, where $\delta_{*,*^\prime}$ is the Kronecker delta. Note that Eq.~\eqref{eq: raw_p_val_basis} uses single particle wavefunctions, an approximation discussed in Sec.~\ref{sec: pO_intro}.

The well confinement energy $U_{j,k,l}$ in Eq.~\eqref{eq: U_cell} is time-reversal symmetric, guaranteeing that its projection into the $p$-orbital-valley Hilbert space has the same symmetry. To represent Eq.~\eqref{eq: U_disorder} as a 2D matrix, we choose the basis ordering $\{\ket{p_+v_{+k_0}},\ket{p_+v_{-k_0}},\ket{p_-v_{+k_0}},\ket{p_-v_{-k_0}}\}$ such that the generators $\sigma_z$ and $\tau_z$ are excluded because they split the time-reversal-symmetric states $\ket{p_\pm}$ and $\ket{v_{\pm k_0}}$. Given this representation, Eq.~\eqref{eq: U_disorder} separates into
\begin{equation}\label{eq: U_disorder_mat}
    \delta U_{m,n}^{m^\prime,n^\prime}=H_\epsilon + H_v+H_\text{voc},
\end{equation}
where
\begin{equation}\label{eq: H_eps_dis}
    H_\epsilon\equiv\Re[\delta U_{+,+}^{-,+}]\sigma_x I-\Im[\delta U_{+,+}^{-,+}]\sigma_y I,
\end{equation}
\begin{equation}\label{eq: H_v_dis}
    H_v\equiv\Re[\delta U_{+,+}^{+,-}] I\tau_x-\Im[\delta U_{+,+}^{+,-}] I\tau_y,
\end{equation}
and
\begin{equation}\label{eq: HVOC}
    \begin{aligned}
        &H_\text{voc} \equiv \frac{1}{2} \big(\Re[\delta U_{+,+}^{-,-} + \delta U_{+,-}^{-,+}] \sigma_x \tau_x + \\ &\Im[\delta U_{+,-}^{-,+} - \delta U_{+,+}^{-,-}]\sigma_x\tau_y -\Im[\delta U_{+,+}^{-,-} + \delta U_{+,-}^{-,+}] \sigma_y \tau_x\\ & + \Re[\delta U_{+,-}^{-,+} -\delta U_{+,+}^{-,-}]\sigma_y \tau_y\big)\\
        &= \eta_{xx} \sigma_x \tau_x + \eta_{xy} \sigma_x \tau_y + \eta_{yx} \sigma_y \tau_x + \eta_{yy} \sigma_y \tau_y,
    \end{aligned}
\end{equation}
where Eq.~\eqref{eq: H_eps_dis} is a disorder-induced addition to the $p$-orbital electric control Hamiltonian $H_\epsilon$ in Eq.~\eqref{eq: pO_H}, Eq.~\eqref{eq: H_v_dis} is the disorder-induced valley Hamiltonian, and $H_\text{voc}$ is the Hamiltonian coupling the valley and $p$-orbital subspaces. The covariance between any two matrix elements of Eq.~\eqref{eq: U_disorder} is given by
\begin{equation}\label{eq: cov}
\begin{aligned}
    &\overline{\delta U_{m,n}^{m^\prime,n^\prime} \delta U_{p,q}^{p^\prime,q^\prime}}=V_c^2\braket{\Psi_{m,n}|\textbf{r}_{j,k,l}}\braket{\textbf{r}_{j,k,l}|\Psi_{m^\prime,n^\prime}}\\
    &\times\overline{\delta U_{j,k,l}\delta U_{j^\prime,k^\prime,l^\prime}}\braket{\Psi_{p,q}|\textbf{r}_{j^\prime,k^\prime,l^\prime}}\braket{\textbf{r}_{j^\prime,k^\prime,l^\prime}|\Psi_{p^\prime,q^\prime}}\\
    &=V_c^2\overline{\delta U_{j,k,l}^2}\braket{\Psi_{m,n}\Psi_{p,q}|\textbf{r}_{j,k,l}}\braket{\textbf{r}_{j,k,l}|\Psi_{p^\prime,q^\prime}\Psi_{m^\prime,n^\prime}},
\end{aligned}
\end{equation}
where the last equality comes from the independence of each cell's stochastic energy in Eq.~\eqref{eq: U_cell}.

We now numerically calculate the moments of the generator coefficients in Eq.~\eqref{eq: U_disorder_mat} using the device parameters in Table~\ref{Tab: device_params} and with discretized cells $C_{j,k,l}$ of dimension $a_0(1,1,1/4)$ such that $N_\text{c}=2$. We start by calculating a realistic envelope function $\phi_0(z_l)$ for the valley states in Eq.~\eqref{eq: raw_p_val_basis} using the deterministic vertical confinement
\begin{equation}\label{eq: well_profile}
    \tilde{U}_{j,k,l} = e F z_l + \overline{U_{j,k,l}},
\end{equation}
where the vertical electric field $F$ pushes the valence electron wavefunction against the interface to cause larger valley splitting and VOC. We find that all eight generator coefficients in Eq.~\eqref{eq: U_disorder_mat} are approximately independent with zero mean and standard deviations
\begin{subequations}\label{eq: Hpv_STD}
\begin{align}
    \label{eq: Hpv_STD_a} &\sigma_\text{voc}\equiv\text{std}[\eta_{**}] \approx 4~\text{GHz},\\
    \label{eq: Hpv_STD_b} &\sigma_v\equiv\text{std}\left[\frac{1}{4}\text{Tr}[H_v^{\text{disorder}}\cdot I\tau_{x,y}]\right]\approx 6~\text{GHz},\\
   \label{eq: Hpv_STD_c} &\sigma_\epsilon\equiv \text{std}\left[\frac{1}{4}\text{Tr}[H_\epsilon^{\text{disorder}}\cdot \sigma_{x,y}I]\right]\approx 6~\text{GHz}.
\end{align}
\end{subequations}
Direct calculation of the values in Eq.~\eqref{eq: Hpv_STD} can be found in the code repository \cite{caporalettiZenodo2026}. When the interface is wide, many cells of finite variance are sampled and therefore, according to the central limit theorem, each generator coefficient in Eq.~\eqref{eq: U_disorder_mat} is approximately a Gaussian random variable~\cite{losert_PracticalStrategiesEnhancing_2023}. Additionally, a wide interface produces negligible deterministic valley splitting. It follows that the valley splitting is governed by a Rayleigh distribution
\begin{equation}\label{eq: valley_split_dist}
    E_v \sim f_\text{Ray}(E_v|2\sigma_v),
\end{equation}
as derived in Ref.~\cite{losert_PracticalStrategiesEnhancing_2023}, the valley phase by a uniform distribution $\mathcal{U}(0,2\pi)$, and the magnitude of VOC by a Chi distribution
\begin{equation}\label{eq: VOC_Chi}
\begin{aligned}
    E_\text{voc}&\equiv\frac{1}{2}\sqrt{\text{Tr}[H_\text{voc}^2]}=\sqrt{\sum_{i,j}\eta_{i,j}^2}\\ 
    &\sim\frac{1}{\sigma_\text{voc}}\chi\left(\frac{E_\text{voc}}{\sigma_\text{voc}}\bigg|4\right) = \frac{E_\text{voc}^3}{2\sigma_\text{voc}^4}\exp\left({\frac{-E_\text{voc}^2}{2 \sigma_\text{voc}^2}}\right),
\end{aligned}
\end{equation}
where $i,j \in\{x,y\}$ and $\chi(x|k)$ is the Chi distribution of variable $x$ with $k$ degrees of freedom. Using the distributions above, we obtain the expected strengths of VOC and valley splitting:
\begin{subequations}\label{eq: mean_Ham}
\begin{align}
    \label{eq: mean_Ev}&\overline{E_v} = \sqrt{\frac{\pi}{2}}(2\sigma_v) \approx 60~\mu\text{eV} \sim 15~\text{GHz},\\
    &\label{eq: mean_VOC}\overline{E_\text{voc}}=\frac{3}{2}\sqrt{\frac{\pi}{2}}\sigma_\text{voc}\approx 32~\mu\text{eV}\sim 8~\text{GHz}.
\end{align}
\end{subequations}

The full Hamiltonian governing the joint $p$-orbital--valley subspace is given by
\begin{equation}\label{eq: H_full}
    H=H_{pO}+H_\epsilon+H_v+H_\text{voc}
\end{equation}
where we regard $H_\epsilon$ as a static offset that can be compensated for using the control Hamiltonian $H_{pO}$ and therefore left out of Eq.~\eqref{eq: H_main}. In Sec.~\ref{sec: swt_spt_in_poV_spec}, we characterize the spectrum of Eq.~\eqref{eq: H_full} in various parameter regimes, finding that different qubit encodings in the 4-level system arise with sweet spots that are first-order insensitive to fluctuations in the dot anisotropy and orientation.

Although Eq.~\eqref{eq: Hpv_STD} predicts valley splitting and VOC of the same order, one could argue that the extremal regimes $\overline{E_\text{voc}}\ll \overline{E_v}$ and $\overline{E_\text{voc}}\gg \overline{E_v}$ may exist in practice, for the following reasons. First, the VOC theory presented in this section should be considered only a rough estimate, because we are not using the full five-electron ground and first excited state as our ``$p$-orbital'' basis for Eq.~\eqref{eq: raw_p_val_basis}. Additionally, the discretized effective-mass theory of Ref.~\cite{losert_PracticalStrategiesEnhancing_2023} used here is only a phenomenological model. In this sense, the result of Eq.~\eqref{eq: Hpv_STD} primarily demonstrates that VOC is a non-negligible effect that should be considered when operating the $pO$ qubit. One can also imagine adjusting the device parameters in Table~\ref{Tab: device_params} to yield either extremal regime. For example, $\overline{E_\text{voc}}\ll \overline{E_v}$ can be achieved by sharpening the substrate-to-well transition such that the valley splitting becomes deterministically large while VOC, which originates from sampling disorder at the interface, shrinks. One could also consider an oscillating disorder concentration in the well with wavelength similar to that of valley oscillations (wiggle well~\cite{mcjunkinSiGeQuantumWells2022}), again boosting the deterministic valley splitting beyond VOC. Finally, $\overline{E_\text{voc}}\gg \overline{E_v}$ can occur in the presence of interface steps, which reduce the valley splitting~\cite{losert_PracticalStrategiesEnhancing_2023} but enhance VOC as the $p$-orbitals sample distinct interfaces.

If VOC and the valley splitting are truly of the same order, the results of Sec.~\ref{sec: swt_spt_in_poV_spec} become less accurate because VOC heavily mixes all four levels, making the states in Eq.~\eqref{eq: pOv_states} and Eq.~\eqref{eq: pO_v-_states} poorly isolated from the other two levels most of the time. Although a qubit encoding that is well isolated from the others is almost always possible in this case, when VOC $\sim10~\text{GHz}$ the avoided crossings are too large to drive across directly, due to AWG hardware bandwidth limitations. This means that one often has to move away from avoided crossings when encoding a qubit, losing sweet-spot-like dephasing times.

\subsection{Decoherence Calculations}\label{App: sub_Decoherence_Calculation}
Dephasing times are estimated using a phenomenological model for charge noise that we develop in App.~\ref{subApp: chrg_noise_mod}. This model is then applied to both the $pO$-qubit in App.~\ref{subApp: B_z swt spt dephasing} and the $pOv$-qubit in App.~\ref{subApp: pOv_dephasing}.

\subsubsection{Charge Noise Model}\label{subApp: chrg_noise_mod}
Physically, charge noise perturbs the confinement potential in Eq.~\eqref{eq: H_eps}, altering both the dot's anisotropy $\Omega$ and semi-major axis orientation $\alpha$. We model this phenomenon using an ensemble of electric dipoles~\cite{kuhlmann_ChargeNoiseSpin_2013} located at, and oriented parallel to, the interface between the heterostructure and gate electrodes ~\cite{connors_LowfrequencyChargeNoise_2019}. Each dipole contributes an electric potential in the plane of the dot given by
\begin{equation}\label{Eq: single_dip_pot}
    V_i(\textbf{r},\textbf{r}_i,\phi_i)=\frac{e d}{4\pi \epsilon_0 \epsilon_r}\frac{(x-x_i)\cos(\phi_i)+(y-y_i)\sin(\phi_i)}{\left((x-x_i)^2+(y-y_i)^2+D^2\right)^{3/2}},
\end{equation}
where $e$ is the charge of an electron, $d$ is the electric dipole moment length, $\epsilon_0$ is the permittivity of free space, $\epsilon_r$ is the relative permittivity of Si, $D$ is the dot-to-interface distance, and $(x_i,y_i,\phi_i)$ are the center-of-charge coordinates and orientation of the dipole in the interface plane, respectively. We approximate Eq.~\eqref{Eq: single_dip_pot} through a second-order Taylor expansion:
\begin{equation}\label{Eq: dip_pot_Taylor_series}
\begin{aligned}
    V_i&\approx \frac{1}{2}\partial_x^2V_i|_{0} x^2+\frac{1}{2}\partial_y^2V_i|_{0} y^2+\partial_x \partial_y V_i|_{0} xy,
\end{aligned}
\end{equation}
where $|_0$ indicates that the dipole potential is evaluated at the origin (which coincides with the quantum dot center), and the subscript $i$ indicates the dipole from which the term originates and should be taken to imply dependence on $\textbf{r}_i$. In Eq.~\eqref{Eq: dip_pot_Taylor_series}, terms of zero or odd degree are neglected because they produce a global shift in the $p$-orbital energy. Even terms with degree $a>2$ are also neglected because their effect is suppressed by $\sim (l_0/D)^{a-2}$ relative to the degree-$2$ term. Writing the interaction energy resulting from Eq.~\eqref{Eq: dip_pot_Taylor_series} in the basis of Eq.~\eqref{eq: pO_basis}, one obtains
\begin{equation}\label{eq: dip_pot_pO_basis}
\begin{aligned}
    e V_i=&-\frac{e l_0^2}{2}\big[\frac{1}{2}\left(\partial_x^2V_i|_{0}-\partial_y^2V_i|_{0}\right)\sigma_x + \partial_x \partial_y V_i|_{0}\sigma_y\big]\\
    \equiv&~\delta_{x_i}\sigma_x+\delta_{y_i}\sigma_y.
\end{aligned}
\end{equation}
Subsequently, Eq.~\eqref{eq: dip_pot_pO_basis} will be thought of as a vector whose basis vectors are the Pauli matrices $\sigma_{x,y}$.

Generally, one should consider the positions and orientations of the dipoles in Eq.~\eqref{Eq: single_dip_pot} to be time-dependent. However, charge noise in SiGe devices has a $1/f$-like power spectral density~\cite{takeda_CharacterizationSuppressionLowfrequency_2013,freeman_ComparisonLowFrequency_2016,connors_LowfrequencyChargeNoise_2019, connors_ChargenoiseSpectroscopySi_2022} that is largest at low frequency, while the $pO$ qubit is controlled at frequencies $1$--$10$ GHz. We therefore take a quasistatic approximation where Eq.~\eqref{Eq: single_dip_pot} is time-independent. Between qubit initializations, one can imagine that the dipole shifts in location and orientation. In the simplest treatment, $(x_i,y_i)$ are uniformly distributed over $\{-\sqrt{A}/2,\sqrt{A}/2\}$, where $A$ is the dipole position domain area, and $\phi_i$ is uniformly distributed over $\{0,2\pi\}$. In this case, the statistics of Eq.~\eqref{eq: dip_pot_pO_basis} are given by
\begin{equation}\label{eq: max_c_var}
\begin{aligned}
    &\pmb{\mu}=0,\\
    &\pmb{\Sigma}(A) = \delta_{j,k}~g(A),
\end{aligned}
\end{equation}
where $\pmb{\mu}$ is a vector containing the mean of each component of Eq.~\eqref{eq: dip_pot_pO_basis}, $\pmb{\Sigma}(A)$ is the covariance matrix of the components of Eq.~\eqref{eq: dip_pot_pO_basis}, and $g(A)$ is a function of $A$.

Adding together the effect of many dipoles in Eq.~\eqref{eq: dip_pot_pO_basis} simultaneously, we obtain the total perturbation to the confinement potential. Assuming different dipoles are independent and identically distributed, the central limit theorem tells us that when $n$ is large,
\begin{equation}\label{eq: CLT}
    \lim_{n\rightarrow \infty}H_p(n) = \mathcal{N}(0,n\pmb{\Sigma}(A)),
\end{equation}
where $\mathcal{N}(\pmb{\mu},\pmb{\Sigma})$ is a bivariate normal vector. This limit can be taken exactly so long as we hold the dipole density $\rho=n/A$ constant, i.e.\ the device area $A$ extends to infinity. In this limit, Eq.~\eqref{eq: CLT} becomes
\begin{equation}\label{eq: therm_lim_H_p}
    \tilde{H}_p=\mathcal{N}(0,\pmb{\Sigma}_\rho) = \delta_x \sigma_x +\delta_y \sigma_y,
\end{equation}
where
\begin{equation}\label{eq: therm_lim_cov}
    \pmb{\Sigma}_\rho = \lim_{A\rightarrow \infty}\rho A\pmb{\Sigma}(A) = \delta_{j,k} \frac{15\pi}{256}\left(\frac{ed}{4\pi \epsilon_0 \epsilon_r}\right)^2\frac{\rho e^2 l_0^4}{D^6}\equiv\delta_{j,k}\sigma^2,
\end{equation}
$l_0$ is the characteristic dot length, and $D$ is the distance from the dot to the electrodes. Equation~\eqref{eq: therm_lim_H_p} is a bivariate normal vector with zero mean and diagonal covariance matrix $\pmb{\Sigma}_\rho$, and therefore has joint probability density function (PDF)
\begin{equation}\label{eq: noise_pdf_cart}
    f(\delta_x,\delta_y|\sigma)=\frac{1}{2\pi \sigma^2}e^{-\frac{\delta_x^2+\delta_y^2}{2\sigma^2}}.
\end{equation}

The charge-noise Hamiltonian in Eq.~\eqref{eq: therm_lim_H_p} can be added to a deterministic, electrical control Hamiltonian $H_\Lambda$ to obtain
\begin{equation}\label{eq: ctrl_noise}
    H_{\Lambda} + \tilde{H}_p=\left(\Lambda_x+\delta_x\right)\sigma_x + \left(\Lambda_y+\delta_y\right)\sigma_y,
\end{equation}
whose components follow the distribution
\begin{equation}\label{eq: cart_ctrl_noise_PDF}
    f(\Omega_x,\Omega_y|\Lambda_x,\Lambda_y,\sigma)=\frac{1}{2\pi \sigma^2}e^{-\frac{\left(\Omega_x-\Lambda_x\right)^2+\left(\Omega_y-\Lambda_y\right)^2}{2\sigma^2}},
\end{equation}
obtained by changing variables in Eq.~\eqref{eq: noise_pdf_cart} to $\Omega_{x,y}\equiv\Lambda_{x,y}+\delta_{x,y}$. Eq.~\eqref{eq: cart_ctrl_noise_PDF} can be converted to polar coordinates by substituting $\Omega_x = \frac{\Omega}{2} \cos(2\alpha)$, $\Omega_y = \frac{\Omega}{2} \sin(2\alpha)$, $\Lambda_x=\frac{\Omega_0}{2}\cos(2\alpha_0)$, and $\Lambda_y=\frac{\Omega_0}{2}\sin(2\alpha_0)$ to obtain
\begin{equation}\label{eq: pre_Rice_dist}
    f(\Omega,\alpha|\Omega_0,\alpha_0,\sigma)=\frac{\Omega}{4\pi \sigma^2}e^{-\frac{\Omega^2+\Omega_0^2}{8\sigma^2}}e^{\frac{\Omega\Omega_0}{4\sigma^2}\cos(2\alpha_0-2\alpha)}.
\end{equation}
Concretely, Eq.~\eqref{eq: pre_Rice_dist} is the probability density for finding the qubit at $(\Omega,\alpha)$ when attempting to park at $(\Omega_0,\alpha_0)$ while subjected to electrical noise of strength $\sigma$.
Typical noise strengths are of order
\begin{equation}\label{eq: Ray_mom}
\begin{aligned}
    \frac{\sigma}{h} \approx 4~\text{MHz}
\end{aligned}
\end{equation}
which we obtain using the device parameters in Table~\ref{Tab: device_params}.

\subsubsection{$pO$ qubit Dephasing with Out-of-Plane Magnetic Field}\label{subApp: B_z swt spt dephasing}
\begin{figure}[h]
    \centering
    \includegraphics[trim={10cm 3cm 10cm 3cm}, clip, width=.8\columnwidth]{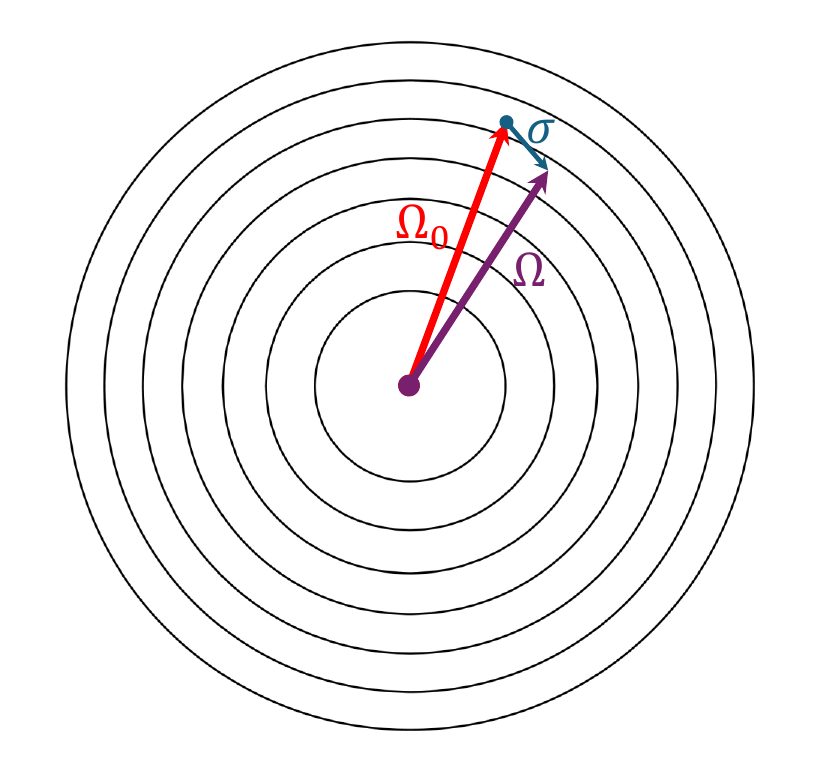}
    \caption{Contour plot of the energy landscape with out-of-plane magnetic field $B_z$. The red vector is the electric control vector with magnitude $\Omega_0$, the blue vector is the stochastic noise vector of characteristic length $\sigma$ with uniformly distributed angle, and the purple vector is their sum, with magnitude $\Omega$. The distribution of $\Omega$ is given by the Rice distribution in Eq.~\eqref{eq: Rice}.\label{fig: B_z_swt_spt}}
\end{figure}
In the absence of noise, a $pO$ qubit in an elliptical dot with an out-of-plane magnetic field $B_z$ has a splitting of
\begin{equation}\label{eq: pO_split_OB}
    E=\sqrt{\Omega^2 + E_z^2}.
\end{equation}
For a Si/SiGe heterostructure in the presence of noise, the dot anisotropy parameters $\Omega$ and $\alpha$ in Eq.~\eqref{eq: pO_H} have a joint probability density function (PDF) $f(\Omega,\alpha|\Omega_0,\alpha_0,\sigma)$ that is derived in Eq.~\eqref{eq: pre_Rice_dist} of App.~\ref{subApp: chrg_noise_mod}, where $\Omega_0$ and $\alpha_0$ are the parameters in the absence of noise, and $\sigma$ is the standard deviation of charge-noise coupling along an arbitrary axis in the $pO$ qubit Bloch sphere equator, derived in App.~\ref{subApp: chrg_noise_mod} to be $4~\text{MHz}$ for a typical scenario.

Integrating $f(\Omega,\alpha|\Omega_0,\alpha_0,\sigma)$ over $\alpha$, we obtain the marginal distribution of $\Omega$:
\begin{multline}\label{eq: Rice}
    f(\Omega|\Omega_0,\alpha_0,\sigma)=\int_{-\frac{\pi}{2}}^{\frac{\pi}{2}}d\alpha~f(\Omega,\alpha|\Omega_0,\alpha_0,\sigma)\\ 
    =\frac{\Omega}{4\sigma^2}e^{-\frac{\Omega^2+\Omega_0^2}{8\sigma^2}}I_0\left(\frac{\Omega \Omega_0}{4\sigma^2}\right) \equiv f_{\text{Rice}}(\Omega|\Omega_0,2\sigma),
\end{multline}
where $I_0(x)$ is the zeroth-order modified Bessel function of the first kind, and $f_{\text{Rice}}(x|\nu,\sigma)$ is the Rice distribution with ``center'' parameter $\nu$ and ``width'' parameter $\sigma$.

The point in parameter space $\Omega_0=0$ satisfies $\partial_\Omega E|_{\Omega_0} = 0$, making it an electrical `sweet spot' because Eq.~\eqref{eq: pO_split_OB} is first-order insensitive to electrical fluctuations. At the sweet spot, Eq.~\eqref{eq: Rice} simplifies to the Rayleigh distribution
\begin{equation}\label{eq: Ray}
    f_{\text{Rice}}(\Omega|0,2\sigma)=\frac{\Omega}{4\sigma^2}e^{-\frac{\Omega^2}{8\sigma^2}} \equiv f_\text{Ray}(\Omega|2\sigma).
\end{equation}
Changing variables in Eq.~\eqref{eq: Ray} from $\Omega$ to $E$, we obtain the sweet spot PDF of Eq.~\eqref{eq: pO_split_OB}:
\begin{equation}\label{eq: swt_spt_PDF}
    f_0(E|E_z,2\sigma)\equiv\frac{E}{4\sigma^2}e^{-\frac{E^2-E_z^2}{8\sigma^2}}.
\end{equation}
Using Eq.~\eqref{eq: swt_spt_PDF}, one can calculate the standard deviation of the energy splitting, $s_E(\Omega_0,2\sigma,E_z)$, at the sweet spot to be
\begin{equation}\label{eq: Bz_std}
    s_E(0,\sigma,E_z)= \frac{4\sigma^2}{E_z} +\mathcal{O}\left(\frac{\sigma^4}{E_z^3}\right).
\end{equation}
For $\sigma/h \approx 4~\text{MHz}$ and $B_z = 20~\text{mT} \rightarrow E_z/h \approx 3~\text{GHz}$, one obtains $s_E(0,\sigma,E_z)=20~\text{kHz}$ and a sweet spot dephasing time
\begin{equation}\label{eq: T_2_@_B_swt_spt}
    T_{2}^{*(\text{electric})}(0)=\frac{\sqrt{2}\hbar}{s_E(0,2\sigma,E_z)}\approx 10~\mu\text{s}.
\end{equation}

One must also consider the effect of magnetic field fluctuations. In Si, the prevalent source of magnetic noise is nuclear spin fluctuations of $^{29}\text{Si}$ isotopic impurities~\cite{witzelQuantumTheoryElectron2006}. The dephasing of the $pO$ qubit due to this noise source can be related to the dephasing of a Loss-DiVincenzo (LD) spin qubit~\cite{loss_QuantumComputationQuantum_1998} due to the same source by equating the ratio of the $pO$ qubit and spin qubit magnetic couplings to the ratio of the $pO$ qubit and Loss-DiVincenzo dephasing times, yielding the relation
\begin{equation}\label{eq: T_2_mag}
    T_2^{*(\text{magnetic})}=\frac{g \mu_B B_z}{E_z} T_2^{*(LD)}=\frac{g m^*}{2 m_e} T_2^{*(LD)} \approx 10~\mu\text{s},
\end{equation}
where the numerical value at the end uses an observed Loss-DiVincenzo dephasing time in isotopically purified Si of $T_2^{*(LD)} = 33~\mu\text{s}$~\cite{chan_AssessmentSiliconQuantum_2018}. Note that the magnetic dephasing time in Eq.~\eqref{eq: T_2_mag} is similar to the electrical dephasing time of Eq.~\eqref{eq: T_2_@_B_swt_spt}, which uses $B_z=20~\text{mT}$. Therefore, increasing $B_z$ further to reduce electrical dephasing would do little to improve the total $pO$ qubit dephasing time, since magnetic dephasing would remain the limiting factor.

Baseband control requires the qubit to have multiple operating points corresponding to different Bloch-sphere rotation axes, so one cannot always remain at the sweet spot. A typical baseband control scheme is to (1) park at the sweet spot for rotations around the $z$-axis of the Bloch sphere and (2) deform the dot to $\Omega_0 \neq 0$ for rotations around an independent axis $\hat{n}$. Linearizing the splitting about an operating point $\Omega_0$, the standard deviation of $E$, $s_E(\Omega_0,2\sigma,E_z) \approx \partial_\Omega E|_{\Omega_0}\, s_\Omega(\Omega_0,2\sigma)$, yields a dephasing time
\begin{align}\label{eq: lin_T_2}
    T_{2}^{*(\text{electric})}(\Omega_0) &\equiv \frac{\sqrt{2}\hbar}{s_E(\Omega_0,2\sigma,E_z)} \nonumber \\
    &\approx \frac{\sqrt{2}\hbar }{s_\Omega(\Omega_0,2\sigma)}\sqrt{1+\left(\frac{E_z}{\Omega_0}\right)^2}.
\end{align}
The percent error in using the linear approximation is on the order of the ratio between the second- and first-order Taylor expansion terms of Eq.~\eqref{eq: pO_split_OB} about $\Omega_0$, evaluated at $s_\Omega(\Omega_0,2\sigma)+\Omega_0$:
\begin{equation}
    \frac{1}{2}\frac{\partial^2_\Omega E}{\partial_\Omega E}s_\Omega(\Omega_0,2\sigma)= \frac{s_\Omega(\Omega_0,2\sigma)}{2\Omega_0}\left(\frac{\Omega_0^2}{E_z^2}+1\right)^{-1}.
\end{equation}
For typical parameters $\sigma/h \approx 4~\text{MHz}$ and $E_z/h \approx 3~\text{GHz}$, the error is $<1\%$ when $\Omega_0/h > 0.4~\text{GHz}$.

To perform any local unitary in a maximum of 5 rotations, the angle between $\hat{z}$ and $\hat{n}$ must be at least $\pi/4$~\cite{hamada_minimumnumberrotations_2014}, so we consider an operating point of $\Omega_0=E_z \approx 3~\text{GHz}$, where Eq.~\eqref{eq: lin_T_2} gives $T_{2}^{*(\text{electric})}(\Omega_0) \approx 40~\text{ns}$. Thus, by pulsing between the sweet spot and $\Omega_0 = E_z$, the $pO$ qubit retains sweet-spot coherence during $\sigma_z$-axis rotations while paying the usual $\sim 40~\text{ns}$ dephasing cost only during the off-axis rotation.

Finally, we show that leakage from the $pO$ qubit subspace due to spin excitation via SOC is negligible. SOC originates from spatial inhomogeneities of the magnetic field. A global, out-of-plane magnetic field can be made effectively homogeneous over the length scale of a dot~\cite{vahapoglu_Singleelectronspinresonance_2021}, so the remaining potential source of SOC is fluctuating nuclear spins of $^{29}\text{Si}$ isotopic impurities. For isotopically purified Si~\cite{koh_HighfidelityGatesQuantum_2013}, fluctuations in spin splittings are $g \mu_B \Delta B_z \sim 0.1~\text{neV}$ over the length scale of two quantum dots ($\sim 100~\text{nm}$)~\cite{petta_DynamicNuclearPolarization_2008}. Extrapolating linearly, we obtain a spin splitting gradient of $\partial_r E_s \sim 1~\text{peV}/\text{nm}$. For $p$-orbitals roughly separated by a dot size of $l_0 \sim 10~\text{nm}$, we therefore get a SOC strength $\sim 10~\text{peV} \approx 2~\text{kHz}$. Dynamics on this time scale are two orders of magnitude slower than the slowest dephasing rate, $\sim 1/T_{2,0}^{*(\text{electric})} \approx 100~\text{kHz}$, from Eq.~\eqref{eq: T_2_@_B_swt_spt}, and can therefore be neglected.

\subsubsection{$pOv$ qubit Dephasing Time}\label{subApp: pOv_dephasing}
In the presence of VOC, the $pOv$-qubit has an energy splitting
\begin{equation}\label{eq: pOv_split}
\begin{aligned}
    2|H_{pOv}|=2 \sqrt{\Delta(\alpha)^2+\left(\frac{\Omega-E_v}{2}\right)^2},
\end{aligned}
\end{equation}
where $c,d$ are functions of the VOC generator coefficents and defined in Eq.~\eqref{eq: H_pOv}. Calculating the standard deviation of Eq.~\eqref{eq: pOv_split}, required for estimating $T_2^*$, becomes analytically tractable if we approximate the splitting using a truncated, second-order Taylor series given by
\begin{equation}\label{eq: pOv_split_series}
\begin{aligned}
    2|H_{pOv}| \approx \bigg(&2|H_{pOv}| +  2\nabla|H_{pOv}| \cdot d_0 \\
    &+d_0^T \cdot \nabla\nabla|H_{pOv}| \cdot d_0 \bigg)\bigg|_{\Omega_0,\alpha_0},    
\end{aligned}
\end{equation}
where $d_0 \equiv (\Omega-\Omega_0,2\alpha-2\alpha_0)$ and $\nabla\nabla$ is the Hessian operator given by 
\begin{equation}\label{eq: Hess}
    \nabla\nabla  \equiv 
    \begin{pmatrix}
    \partial_\Omega^2
    & \partial_\Omega\partial_\alpha - \dfrac{1}{\Omega}\partial_\alpha \\[7pt]
     \partial_\Omega\partial_\alpha - \dfrac{1}{\Omega}\partial_\alpha
    & \partial_\alpha^2 + \Omega \partial_\Omega
    \end{pmatrix}.
\end{equation}
This approximation works when parked at the sweet spot because second order terms are included. Although the first and second order terms in Eq.~\eqref{eq: pOv_split_series} diverge at true level crossings in the spectrum, causing the dephasing time to go to zero, the actual dephasing time at these crossings is a minimum of $\sim40~\text{ns}$ anyway, and therefore Eq.~\eqref{eq: pOv_split_series} produces a lower bound on $T_2^*$. 

We now restrict our focus to $\Omega_0=E_v$, as this is the relevant $pO$-splitting when at the $pOv$ sweet spot and while performing two-qubit operations. If one deviates from $\Omega_0=E_v$ by $\gtrsim\pm1~\text{GHz}$, the dephasing time approaches the nominal dipole protected dephasing $T_2^*\sim40~\text{ns}$. 

Suppose that the deviation away from $(E_v,2\alpha_0)$, given by $(\Omega,2\alpha)$, originates from the charge noise perturbation in Eq.~\eqref{eq: therm_lim_H_p} which in polar coordinates is given by
\begin{equation}
    \tilde{H}_p=\delta(\cos(\phi)\sigma_x+\sin(\phi)\sigma_y),
\end{equation}
where $\delta$ and $\phi$ follow the distribution
\begin{equation}\label{eq: noise_polar_PDF}
    f(\delta,\phi)=\frac{\delta }{2\pi\sigma^2}e^{-\frac{\delta^2}{2\sigma^2}}.
\end{equation}
Adding $\tilde{H}_p$ to $H_{pO}(E_v,\alpha_0)$ yields the perturbed controls $(\Omega,2\alpha)$ which transform $d_0$ into 
\begin{equation}\label{eq: polar_dif_vec}
\begin{aligned}
    d_0 = \big(&\sqrt{4\delta^2+E_v^2+4 \delta E_v\cos(2\alpha_0-\phi)}-E_v,\\
    &\arctan(E_v \cos(2\alpha_0) + 2\delta\cos(\phi),\\&
    ~~~~~~~~~~E_v \sin(2\alpha_0) + 2\delta\sin(\phi))-2\alpha_0\big).
\end{aligned}
\end{equation}
Further simplification can be made by expanding Eq.~\eqref{eq: pOv_split_series} in terms of the small parameter $\delta/E_v$.
For valley splitting $\overline{E_v}/h\sim 1$--$10~\text{GHz}$ and noise strength $\mathbb{E}[\delta]/h = \frac{1}{h}\sqrt{\frac{\pi}{2}}\sigma \approx 5~\text{MHz}$, we have $\delta/E_v\sim 10^{-3}$--$10^{-4}$. Truncating Eq.~\eqref{eq: pOv_split_series} to zeroth order in $\delta/E_v$, one gets 
\begin{equation}\label{eq: pOv_split_approx}
    2|H_{pOv}|\approx 2|H_{pOv}|_{\Omega_0,\alpha_0} + \frac{\delta^2\cos(2\alpha_0-\phi)^2}{|\Delta(\alpha_0)|}.
\end{equation}
Using the PDF in Eq.~\eqref{eq: noise_polar_PDF}, the standard deviation of the $pOv$ splitting is calculated, yielding a dephasing time
\begin{equation}\label{eq: T_2_gen_analytic}
\begin{aligned}
    T_2^*\approx \frac{\sqrt{2}\hbar}{\text{std}[2|H_{pOv}|]}&=\frac{\hbar|\Delta(\alpha_0)|}{\sigma^2},
\end{aligned}
\end{equation}
where $\text{std}[*]$ indicates the standard deviation over charge-noise realizations and $\Delta(\alpha_0)$ is the strength of single-qubit $X$ rotations in the $pOv$-qubit Hamiltonian or Eq.~\eqref{eq: H_pOv}.

\subsubsection{Noisy Two $pOv$-Qubit Pulse Implementation}\label{subApp: two_pOV_pulse_inf}
To estimate the effect of charge noise on the optimized two-qubit pulse $U$ from Eq.~\eqref{eq: pulse_seg}, we must compute the fluctuation in dot orientations $\alpha$, anisotropies $\Omega$, and interdot distance $L$ due to the potential created by a dipole TLF given in Eq.~\eqref{Eq: single_dip_pot}. The perturbation of $\Omega_{1,2}$ and $\alpha_{1,2}$ due to a dipole TLF comes directly from Eq.~\eqref{eq: dip_pot_pO_basis} with the caveat that the in-plane derivatives in Eq.~\eqref{Eq: dip_pot_Taylor_series} must be about each qubit's center.

We place the two dots at $\mathbf{r}_1=(-L/2,0)$ and $\mathbf{r}_2=(+L/2,0)$ along the interdot axis. Evaluating the second derivatives in Eq.~\eqref{Eq: dip_pot_Taylor_series} about each center $\mathbf{r}_q$ and summing the resulting contributions Eq.~\eqref{eq: dip_pot_pO_basis} over the $n=\rho A$ dipoles of one quasistatic realization gives a dot-dependent noise vector $\pmb{\delta}^{(q)}=(\delta_x^{(q)},\delta_y^{(q)})$, where
\begin{equation}\label{eq: per_dot_noise}
\begin{aligned}
    &\delta_x^{(q)} = -\frac{e l_0^2}{4}\sum_i\big(\partial_x^2 V_i-\partial_y^2 V_i\big)\big|_{\mathbf{r}_q},\\
    &\delta_y^{(q)} = -\frac{e l_0^2}{2}\sum_i \partial_x\partial_y V_i\big|_{\mathbf{r}_q}.    
\end{aligned}
\end{equation}
This is the finite-$n$ realization of Eq.~\eqref{eq: CLT} for two dots immersed in the same dipole field, making $\pmb{\delta}_1$ and $\pmb{\delta}_2$ correlated. Following Eq.~\eqref{eq: ctrl_noise}, $\pmb{\delta}_q$ is added to the deterministic control of qubit $q$, $\pmb{\Lambda}_q=\tfrac{\Omega_0}{2}(\cos 2\alpha_{0,q},\,\sin 2\alpha_{0,q})$, where the nominal anisotropy sits at the sweet spot $\Omega_0=E_v$ for every pulse. Converting the perturbed field $\pmb{\Lambda}_q+\pmb{\delta}_q$ back to polar coordinates yields the fluctuated anisotropy
\begin{equation}\label{eq: Omega_fluc}
    \Omega_q = 2\sqrt{\big(\tfrac{\Omega_0}{2}\cos 2\alpha_{0,q}+\delta_x^{(q)}\big)^2+\big(\tfrac{\Omega_0}{2}\sin 2\alpha_{0,q}+\delta_y^{(q)}\big)^2},
\end{equation}
and orientation
\begin{equation}\label{eq: alpha_fluc}
    \alpha_q = \tfrac12\,\arctan{\!\big(\tfrac{\Omega_0}{2}\cos 2\alpha_{0,q}+\delta_x^{(q)},\tfrac{\Omega_0}{2}\sin 2\alpha_{0,q}+\delta_y^{(q)}}\big). 
\end{equation}

The electric field $F_x^{(q)}$ at qubit $q$ produced by charge noise causes fluctuations in the qubit's lateral position, assuming harmonic confinement. This effect was omitted for the one qubit system in Sec.~\ref{subApp: chrg_noise_mod} because the $p$-orbital splitting is independent of the dot's lateral position. However, for the two qubit system discussed in Sec.~\ref{sec: two qubit}, relative position shifts are equivalent to changes in $L$, ultimately causing fluctuations in the two qubit interaction $\Omega_{zz}$ which goes like $1/L^5$. Given that each dot has a confinement frequency $\omega_0=\hbar/(m l_0^2)$, the shift of qubit $q$ along the $x$-axis connecting the qubits is given by
\begin{equation}\label{eq: dot_disp}
    \delta x_q = \frac{eF_x^{(q)}}{m\omega_0^2}
    = -\frac{e}{m\omega_0^2}\sum_i \partial_x V_i\big|_{\mathbf{r}_q}.
\end{equation}
The interdot distance then fluctuates by the relative displacement of the two dots,
\begin{equation}\label{eq: L_fluc}
    L \rightarrow L+\delta L,\qquad \delta L = \delta x_2-\delta x_1 .
\end{equation}

For a given quasistatic realization, every pulse segment $k$ of the optimized sequence inherits the same $\pmb{\delta}_{1,2}$ and $\delta L$, but since each pulse parks at a different nominal orientation, the resulting $(\Omega_{q,k},\alpha_{q,k})$ from Eqs.~\eqref{eq: Omega_fluc}-\eqref{eq: alpha_fluc} differ segment to segment. Substituting $\Omega_{q,k}$, $\alpha_{q,k}$, and $L+\delta L$ into $U$ from Eq.~\eqref{eq: pulse_seg} creates the noisy gate $\tilde{U}_{2q}$. Note that although $\Omega$ does not appear directly in Eq.~\eqref{eq: pulse_seg} because setting $\Omega_0=E_v$ zeros the single qubit term $(\Omega_0-E_v)/2 ~\sigma_z^{(q)}$, fluctuations in $\Omega$ still impact the infidelity of $\tilde{U}_{2q}$ because they cause this single-qubit term to deviate from zero.

For the charge noise simulations in Sec.~\ref{sec: two qubit}, we use the charge noise parameters from Table~\ref{Tab: device_params} along with a finite heterostructure-electrode interface area $A=10^4~\text{nm}^2$. Given this area and the dipole density in Table~\ref{Tab: device_params}, a charge noise configuration has $10$ dipole TLFs. The code used to produce the results in Fig.~\ref{fig: inf_plot} can be found in the repository~\cite{caporalettiZenodo2026}.
%%%%%%%%%%%%%%%
\bibliography{VOC_ref}

\end{document}